\documentclass[10pt,psfig]{article}
\pdfoutput=1
\usepackage[square,sort,comma,numbers]{natbib}
 \usepackage{amsmath}
 \usepackage{amssymb}
\usepackage{color}
\usepackage{placeins}
\usepackage{ulem}
\usepackage{graphicx}
\usepackage{leftidx}

\usepackage{xcolor,colortbl}
\definecolor{Gray}{gray}{0.85}
\newcolumntype{g}{>{\columncolor{Gray}}c}
\newcolumntype{w}{>{\columncolor{white}}c}
\definecolor{LightCyan}{rgb}{0.88,1,1}

\newcommand{\tb}{\textcolor{black}}

\newcommand{\tc}{\textcolor{black}}

\newcommand{\tr}{\textcolor{black}}

\newcommand{\tor}{\textcolor{black}}

\newcommand{\tbl}{\textcolor{black}}

\newcommand{\Var}{{\rm Var\,}}

\newcommand{\var}{\Var}

\newtheorem{Remark}{Remark}[section]

\newtheorem{Algorithm}{Algorithm}[section]

\usepackage{scrextend}
\usepackage{pslatex}
\usepackage{graphicx}

\makeatletter
\gdef\@ptsize{2}% 12pt documents
\let\@currsize\normalsize
\makeatother
\usepackage{setspace}
\doublespace
\usepackage[font=small, width=0.9\textwidth]{caption}

\usepackage{microtype}  % avoids citations that hang into the margin

\usepackage{footnote}
\makesavenoteenv{tabular}
\makesavenoteenv{table}

\usepackage{rotating}  % Enables the sideways environment (NCPW)
\usepackage{array}  % Enables "m" tabular environment http://ctan.org/pkg/array
\usepackage{booktabs}  % Enables \toprule  http://ctan.org/pkg/array

 \title{Model-Based and Model-Free point prediction algorithms for locally stationary random fields}

\author {Srinjoy Das \\
School of Mathematical and Data Sciences \\
West Virginia University \\
Morgantown, WV, 26506 USA \\
email: {\tt   srinjoy.das@mail.wvu.edu}
\and
Yiwen Zhang \\
MIT Sloan School of Management\\
Cambridge, MA 02142 \\
email: {\tt yiz789@mit.edu}
\and
Dimitris N. Politis \\
Department of Mathematics and \\
Halicioglu Data Science Institute \\
   University of California---San Diego \\
   La Jolla, CA 92093, USA \\
   email: {\tt dpolitis@ucsd.edu}
}
\date{ }

\begin{document}

\newcolumntype{g}{>{\columncolor{Gray}}c}
\maketitle

\newpage

\begin{abstract}

The Model-free Prediction Principle has been successfully applied to general regression problems, as well as problems involving stationary and locally stationary time series. In this paper we demonstrate how Model-Free Prediction can be applied to handle random fields that are only locally stationary, i.e., they can be assumed to be stationary only across a limited part over their entire region of definition. We construct  one-step-ahead point predictors and compare the performance of Model-free to Model-based prediction using models that incorporate a trend and/or heteroscedasticity. Both aspects of the paper, Model-free and Model-based, are novel in the context of random fields that are locally (but not globally) stationary. We  demonstrate the application of our Model-based and Model-free point prediction methods to synthetic data as well as images from the CIFAR-10 dataset and in the latter case show that our best Model-free point prediction results outperform those obtained using Model-based prediction.

%The Model-free Prediction Principle of Politis (2015) has been successfully applied to general regression problems, as well as problems involving stationary time series. However, with long time series, e.g. annual temperature measurements spanning over 100 years or daily financial returns spanning several years, it may be unrealistic to assume stationarity throughout the span of the dataset. In the paper at hand, we show how Model-free Prediction can be applied to handle time series that are only locally stationary, i.e., they can be assumed to be stationary only over short time-windows. Surprisingly there is little literature on point prediction for general locally stationary time series even in model-based setups, and there is no literature whatsoever on the construction of prediction intervals of locally stationary time series. We attempt to fill this gap here as well. Both one-step-ahead point predictors and prediction intervals are constructed, and the performance of model-free is compared to model-based prediction using models that incorporate a trend and/or heteroscedasticity. Both aspects of the paper, model-free and model-based, are novel in the context of time-series that are locally (but not globally) stationary. We also demonstrate the application of our Model-based and Model-free prediction methods to speleothem climate data which exhibits local stationarity and show that our best model-free point prediction results outperform that obtained with the RAMPFIT algorithm previously used for analysis of this type of data.
\end{abstract}


\newpage 

\section{Introduction}

\tb{Consider a real-valued random field dataset $\{Y_{\underline t}, \underline t \in Z^2\}$ defined over a 2-D index-set $D$ e.g. pixel values over an image or satellite data observed on an ocean surface.
% real-valued time series dataset    $Y_1, \ldots, Y_n $ spanning a   long time interval, e.g. annual temperature measurements spanning over 100 years or daily financial returns spanning several years.
 It may be unrealistic to assume that 
% the stochastic structure of time series $\{Y_t , t\in {\bf Z}  \}$ has stayed  invariant over such a  long stretch of time; 
the stochastic structure of such a random field {$Y_{\underline t}$} has stayed  invariant 
% over such a  long stretch of time; 
over the entire region of definition $D$
% hence, we ca nnot assume that $\{Y_t\}$ is stationary. 
hence, we cannot assume that $\{Y_{\underline t}\}$ is stationary.}
Therefore it is more realistic to assume a slowly-changing 
stochastic structure, i.e.,  a {\it locally  stationary model}. Discussions of \tb{such models for locally stationary time series } can be found in
\tr {\cite{priestley1965evolutionary}, \cite{priestley1988non}, \cite{dahlhaus1997fitting} and \cite{dahlhaus2012locally}.}  \tb{In the context of random fields, locally stationary models have been proposed in \cite{kurisu2022nonparametric} and references therein where the data $Y_{\underline t}$ is defined over a continuous subset $S$ of $R^d$. In this paper we assume a locally stationary model for random fields $Y_{\underline t} \in R$ defined over $\underline t \in S$ where $S \subset Z^d$, $d = 2$. Given data $Y_{\underline t_1}, Y_{\underline t_2}, \ldots, Y_{\underline t_n}$, 
% It should be noted that a locally stationary model can also be applicable for random fields which is the subject of this paper. Here our objective is constructing a point predictor
% for the next  data point $Y_{n+1}$, i.e., constructing a point and interval predictor for $Y_{n+1}$.
our objective is to perform \tb{point prediction} for a {\it future} unobserved data point $Y_{\underline t_{n+1}}$. Here $\underline t_1, \underline t_2, \ldots, \underline t_n, \underline t_{n+1} \in Z^2$ denote the coordinates of the random field over the 2-D index set $D$ and the notion of a {\it future} datapoint over a coordinate of a random field for purposes of predictive inference over {$\underline t \in Z^2\ $} is defined in Section \ref{RF.Causality}. Algorithms for point prediction and prediction intervals of  locally stationary time series and their applications in both synthetic and real-life datasets have been discussed in \cite{das2021predictive}. Our work in this paper extends this framework to point prediction over locally stationary random fields with applications involving both synthetic and real-life image data.}

% i.e., constructing a point and interval predictor for $Y_{\underline t_{k}}$.
% A secondary goal is to conduct inference
% regarding parameters of the conditional distribution of $Y_{n+1}$ given the data $\underline{Y}_n=(Y_1,\ldots, Y_n)'$;
%for example, construct point and interval estimates  for
%the conditional expectation $E(Y_{n+1}|\underline{Y}_n)$.

The usual approach  for dealing with nonstationary series is to assume 
that the data can be decomposed as the sum of three components:
$$  
 \mu(\underline t)+ S_{\underline t} + W_{\underline t} 
$$  
where $\mu(\underline t)$ is a deterministic trend function, $S_{\underline t}$ is a seasonal (periodic) 
% time 
series, and $\{W_{\underline t}\}$  is (strictly)  stationary with mean zero;
this is   
the `classical' decomposition of a time series to trend, seasonal and stationary components see e.g. \tr {\cite{brockwell2013time}} which can also be used for decomposition of nonstationary random field data.    The seasonal (periodic) component, be it random or deterministic, can be easily estimated and
removed and 
% see e.g. \tr {\cite{brockwell2013time}}.
% Brockwell and Davis (1991). 
having done that, the `classical' decomposition
simplifies to the following model with additive trend, i.e.,
\begin{equation}
Y_{\underline t}=\mu(\underline t)+   W_{\underline t}
 \label{RF.eq.model homo}
\end{equation}
which 
 can be generalized to accommodate
a     
% time-changing
coordinate-changing 
variance as well, i.e., 
 \begin{equation}
Y_{\underline t}=\mu(\underline t)+ \sigma (\underline t) W_{\underline t} . 
 \label{RF.eq.model hetero}
\end{equation}
In both above models, the %time
series $\{W_{\underline t}\}$
  is assumed to be (strictly)  stationary,
weakly dependent, e.g. strong mixing, and satisfying $EW_{\underline t} =0$;
in model \eqref{RF.eq.model hetero}, it is also
assumed that    $\var(W_{\underline t})=1$.  
As usual, the deterministic functions 
 $\mu( \cdot)$ and  $\sigma (\cdot)$ are unknown but
assumed to belong to 
 a class of functions  that is either finite-dimensional (parametric) or not \tr {(nonparametric)};
we will focus on the latter, in which case
it is customary to assume that $\mu( \cdot)$ and  $\sigma (\cdot)$  possess some degree of smoothness, i.e., that  $\mu(\underline t)$ and  $\sigma (\underline t)$ change
smoothly (and slowly) with $\underline t$.

As far as capturing the  first two moments of $Y_{\underline t}$, 
models \eqref{RF.eq.model homo} and  \eqref{RF.eq.model hetero} 
are considered general and
 flexible---especially when $\mu( \cdot)$ and  $\sigma (\cdot)$ 
are not parametrically specified---and have been studied 
extensively in the case of time series; see e.g. 
\tr {\cite{zhou2009local}, \cite{zhou2010simultaneous}.}
% Zhang and Wu (2011), and Zhou and Wu (2009, 2010).
However,   it may be  that the skewness and/or kurtosis of $Y_{\underline t}$   changes with $\underline t$,
in which case   centering and studentization alone cannot render the problem stationary. To see why, note that  
under model  \eqref{RF.eq.model hetero},
$ EY_{\underline t}=\mu(\underline t)$ and $ \Var Y_{\underline t}=\sigma ^2( \underline t)$; hence, 
\begin{equation}
W_{\underline t} =\frac{Y_{\underline t}-\mu( \underline t)}{ \sigma (\underline t)}      
\label{RF.eq.W}
\end{equation}
  cannot be (strictly) stationary unless 
the skewness and  kurtosis of $Y_{\underline t}$ are constant. 
Furthermore, it may be the case that the nonstationarity is due to 
a feature of the $m$--th dimensional marginal distribution not being constant
for some $m\geq 1$, 
e.g., perhaps the correlation Corr$(Y_{\underline t_j}, Y_{\underline t_{j+1}})$ \tb{where $\underline t_j, \underline t_{j+1} \in Z^2$}
changes smoothly (and slowly) with $\underline t_j$. Notably, models \eqref{RF.eq.model homo} and  \eqref{RF.eq.model hetero} only concern themselves with features of the 
1st marginal distribution.

For all the above reasons, it seems valuable to develop a methodology for
the statistical analysis of nonstationary random fields that does not
rely on  simple additive models such as   \eqref{RF.eq.model homo} and  \eqref{RF.eq.model hetero}. Fortunately, the 
Model-free Prediction Principle of \tb {\cite{Politis2013}, \cite{politis2015model}}
 \tb{suggests a way} to accomplish  Model-free inference
% ---including the  construction of  prediction intervals---
 in the general setting of
   random fields that are only locally stationary.
The key towards Model-free inference is to be able to construct an invertible transformation 
$H_n: \underline{Y}_{\underline t_n} \mapsto \underline \epsilon_{n}$
where $ \underline{Y}_{\underline t_n} = (Y_{\underline t_1}, Y_{\underline t_2}, \ldots, Y_{\underline t_n})$ denotes the random field data under consideration and
$\underline \epsilon_{n}=(\epsilon_{1}, \ldots, \epsilon_{n} )'$
is  a random vector with i.i.d.~components;
the details for point prediction are given in Section \ref{RF.Model-free inference}.
In Section \ref{RF.Model-based inference} we visit the problem of model-based  
inference and develop a point prediction methodology for locally stationary random fields.
% a bootstrap methodology for the construction of  (model-based) prediction intervals. 
Both approaches, Model-based \tb {of Section \ref{RF.Model-based inference}} and Model-free \tb{of Section \ref{RF.Model-free inference}}, are novel, and they are empirically compared to each other in Section \ref{RF.Numerical} \tb {using finite sample \tor{experiments}}. % Both synthetic and real-life data are used for this purpose. 

\graphicspath{{/Users/rumpagiri/Documents/NONPARAMETRIC/model_free/papers/RANDOM_FIELDS}}
\DeclareGraphicsExtensions{.png}

{\begin{figure}[!t]
  \centering
  \includegraphics[width=3.5in, height=2.5in]{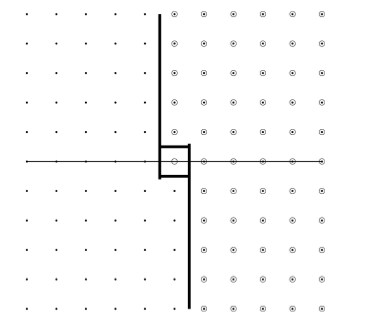}
  \caption{Non Symmetric Half-Plane}
  \label{NSHP}
\end{figure}}

\section{Causality of Random Fields}
\label{RF.Causality}

Given the random field observations {$Y_{\underline t_{1}}, \ldots, Y_{\underline t_{n}}$} our goal is predictive inference for the "next" unknown datapoint {$Y_{\underline t_{n+1}}$}. In this context a definition of causality is necessary to specify the random field coordinate $ \underline t_{n+1}$ where predictive inference will be performed. For this purpose we adopt the framework proposed in \cite{choi2007modeling} and consider random fields discussed in this paper to be defined over a subset of the non symmetric half-plane (NSHP) denoted as $H_{\infty}$. Figure \ref{NSHP} shows an NSHP centered at (0, 0). The NSHP can also be centered at any other point $\underline t$ as follows:
\begin{equation}
NSHP(\underline t) = {\underline t + \underline s \ \ \ \ \forall \underline s \in NSHP(0,0)}
\label{eq.NSHP}
\end{equation}

Such non symmetric half-planes have been used previously for specifying causal 2-D AR models \cite{choi2007modeling}. In such cases a causal 2-D AR model with  $H_p \subset H_{\infty}$ can be defined as below in equation (\ref{eq.2D_AR}) where the set $H_p$ is termed as the region of support (ROS) of the 2-D AR model. . Here $H_p = \{(j,k) \ | \ j = 1,2, \ldots, p \ \ k = 0, \pm 1, \ldots, \pm p\} \cup \{(0,k) \ | \ k = 1,2,\ldots,p\}$ and $v_{t_1, t_2}$ is a 2-D white noise process with mean $0$ and variance $\sigma^2 > 0$. 
\begin{equation}
Y_{t_1, t_2} = \sum \limits_{(j,k) \in H_p} \beta_{j,k} Y_{t_1-j, t_2-k} + v_{t_1, t_2}
\label{eq.2D_AR}
\end{equation}

Based on \cite{dudgeon1984multidimensional} a 2-D AR process with ROS $S$ is causal if there exists a subset $C$ of $Z^2$ satisfying the following conditions:
\begin{itemize}
\item The set C consists of 2 rays emanating from the origin and the points between the rays
\item The angle between the 2 rays is strictly less than 180 degrees
\item $S \subset C$
\end{itemize}

In this case since $H_p \subset H_{\infty}$ satisfies these conditions the 2-D AR process denoted by (\ref{eq.2D_AR}) is causal. Therefore we can use this framework to describe a causal random field defined over the NSHP and perform predictive inference on the same. Given this our setup for point prediction of random fields is described as below.
% construction of both one-step ahead predictors and prediction intervals. This is described below.
%and will be used for defining the "future" point for prediction in both the model-based and model-free inference discussed in this paper. As defined in \cite{dudgeon1984multidimensional} a 2D AR process with an ROS $S$ is causal if there exists a subset $C$ of $Z^2$

Consider random field data $\{Y_{\underline t}, \ \underline t \in E\}$ where $E$ can be any finite subset of $Z^2$ for e.g.
% $E_{\underline n} = \{\underline t \in Z^2$ with $0 < t_1 < n_1 \ \& \ 0 < t_2 < n_2, \ \underline n=(n_1, n_2)\}$. 
$E_{\underline n} = \{\underline t \in Z^2$ with \ $\underline n=(n_1, n_2)\}$. 
Our goal is predictive inference at $\underline t = (t_1, t_2)$ where  $0 < t_1 < n_1 \ \& \ 0 < t_2 < n_2,$. This "future" value $Y_{t_1, t_2}$ is determined using data defined over the region as shown in Figure  \ref{NSHP_pred}:
$$
E_{\underline t, \underline n} = NSHP(\underline t) \cap E_{\underline n}
$$

Both model-based and model-free causal inference for $Y_{t_1, t_2}$ are performed using the data specified over this region $E_{\underline t, \underline n}$. We consider predictive inference at $Y_{\underline t} = Y_{t_1, t_2}$ given the data ($Y_{\underline s} \mid \underline s \prec \underline t  \ \& \ \underline s \in E_{\underline t, \underline n}$) where the symbol $\prec$ denotes lexicographical ordering on the region of support of the random field i.e. $ (a_k, b_k) \prec (a_{k+1}, b_{k+1})$ if and only if either $a_k < a_{k+1}$ or $(a = a_{k+1}$ and $b_k < b_{k+1})$ \cite{choi2007modeling}. 
% In the subsequent discussion this lexicographically ordered "past" data $Y_{\underline t_1}, Y_{\underline t_2}, \ldots, Y_{\underline t_n}$ will be denoted as $Y_{\underline s}$ and point prediction will be performed at $Y_{\underline t} = Y_{\underline t_{n+1}}$.
\tb{In the subsequent discussion is the lexicographically ordered "past" data $Y_{\underline s}$ will be denoted as $Y_{\underline t_1}, Y_{\underline t_2}, \ldots, Y_{\underline t_n}$ and point prediction will be performed at $Y_{\underline t} = Y_{\underline t_{n+1}}$.}

%In the subsequent discussion random field data over the region $E_{\underline t, \underline n}$used for predictive inference will be denoted as $\underline{Y}_{\underline t_n} = ({Y}_{\underline t_1}, \ldots, {Y}_{\underline t_n})$ where $Y_{\underline t_1}$ denotes the data at $\underline t = (0, 0)$. Prediction is performed for $\underline t = (t_1, t_2)$ as shown in Figure \ref{NSHP_pred}, this datapoint will be denoted as $Y_{\underline t_{n+1}}$.

\graphicspath{{/Users/rumpagiri/Documents/NONPARAMETRIC/model_free/papers/RANDOM_FIELDS}}
\DeclareGraphicsExtensions{.png}

{\begin{figure}[!t]
  \centering
  \includegraphics[width=3.5in, height=3.0in]{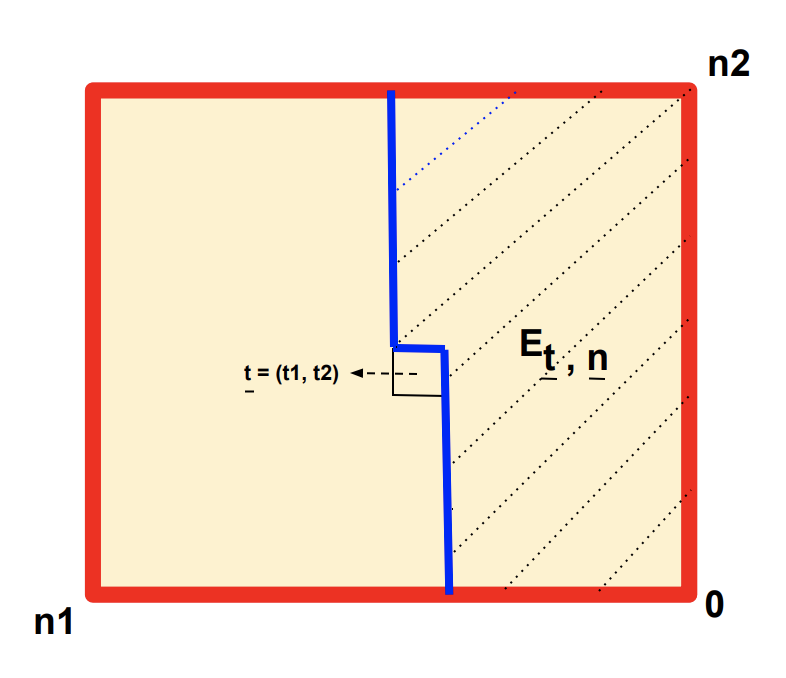}
  \caption{Prediction point for NSHP}
  \label{NSHP_pred}
\end{figure}}

\section{Model-based inference}
\label{RF.Model-based inference}

Throughout Section \ref{RF.Model-based inference}, we will assume  
model   \eqref{RF.eq.model hetero}---that
includes model  \eqref{RF.eq.model homo} as a special case---together with
a nonparametric assumption on smoothness of $\mu(\cdot)$ and  $\sigma ( \cdot)$.
% as described in Remark \ref{RF.re.smooth}.
   
\subsection{Theoretical optimal point prediction}
\label{RF.sec.TOPP}

It is well-known that the $L_2$--optimal 
predictor of  $Y_{\underline t_{n+1}}$ given the data 
$Y_{\underline{s}} = $ %(Y_{\underline t_1},\ldots, Y_{\underline t_n})'$ 
$\underline{Y}_{\underline t_n}=(Y_{\underline t_1},\ldots, Y_{\underline t_n})'$ 
is the conditional expectation $E(Y_{\underline t_{n+1}}|\underline{Y}_{\underline t_n})$ where $\underline{Y}_{\underline t_n}$ indicates the data ${Y}_{\underline t_1}, \ldots, {Y}_{\underline t_n}$.
Furthermore, under model  \eqref{RF.eq.model hetero}, we have 
\begin{equation}
E(Y_{\underline t_{n+1}}|\underline{Y}_{\underline t_{n}})=\mu(\underline t_{n+1})+ \sigma (\underline t_{n+1}) E(W_{\underline t_{n+1}}|\underline{Y}_{\underline t_n}).
\label{RF.eq.point pred}
\end{equation}

% For \tr {$j<J$}, 
%For $\underline t_j \prec \underline t_J$ define ${\mathcal F}_{\underline t_j}^{\underline t_J}(Y)$ to be the
%{\it information set}   $\{ Y_{\underline t_j}, Y_{\underline t_{j+1}},\ldots, Y_{\underline t_J}\}$,
For $\underline j \prec \underline J$ define ${\mathcal F}_{\underline j}^{\underline J}(Y)$ to be the
{\it information set}   $\{ Y_{\underline j}, \ldots, Y_{\underline J}$\},
also known as     $\sigma$--field, 
and note that the information sets 
${\mathcal F}_{-\infty}^{\underline t}(Y)$ and ${\mathcal F}_{-\infty}^{\underline t}(W)$ are identical for any $\underline t$,
i.e., knowledge of  $\{ Y_{\underline s}$ for 
% $\underline s \prec \underline t \ \& \ \underline s \in E_{\underline t, \underline n}$\}
$\underline s \prec \underline t$\}
is equivalent to 
knowledge of { 
% $\{ W_{\underline s}$ for $\underline s \prec \underline t \ \& \ \underline s \in E_{\underline t, \underline n}\}$. 
$\{ W_{\underline s}$ for $\underline s \prec \underline t \}$. 
Here $\mu(\cdot)$ and  $\sigma ( \cdot)$
are assumed known and the symbol $\prec$ denotes lexicographical ordering on the region of support of the random field as described in Section \ref{RF.Causality}. 
Hence, for large $n$, and due to the assumption that $W_{\underline t}$
is weakly dependent (and therefore
  the same must be true for $Y_{\underline t}$ as well), the following large-sample 
approximation is useful, i.e., 
%\begin{equation} 
%E(W_{n+1}|\underline{Y}_n)  \simeq  E(W_{n+1}|Y_s, s\leq n) =
%E(W_{n+1}|W_s, s\leq n)  \simeq E(W_{n+1}|\underline{W}_n )
%\label{NSTS.eq.point pred approx}
%\end{equation}
%\begin{equation}
%E(W_{\underline t_{n+1}}|{Y}_{\underline s}) = E(W_{\underline t_{n+1}}|\underline{Y}_{\underline t_n}) 
%\simeq E(W_{\underline t_{n+1}}|Y_{\underline t_s}, \underline s \preceq \underline n) 
%= E(W_{\underline t_{n+1}}|W_{\underline t_s}, \underline s \preceq \underline n)  \simeq E(W_{\underline t_{n+1}}|\underline W_{\underline t_n } )
%\end{equation}
%\label{RF.eq.point pred approx}
\begin{equation}
E(W_{\underline t_{n+1}}|\underline{Y}_{\underline t_n}) = E(W_{\underline t_{n+1}}|{Y}_{\underline s})
\simeq E(W_{\underline t_{n+1}}|Y_{\underline r}, \underline r \preceq \underline s) 
= E(W_{\underline t_{n+1}}|W_{\underline r}, \underline r \preceq \underline s)  
\simeq  E(W_{\underline t_{n+1}}|{W}_{\underline s})
= E(W_{\underline t_{n+1}}|\underline W_{\underline t_n } )
\end{equation}
\label{RF.eq.point pred approx}
where $\underline{W}_{\underline t_n} = (W_{t_1}, \ldots, W_{t_n})'$.

\tb{We therefore need to} construct an approximation for 
% $E(W_{n+1}|\underline{W}_n )$. 
$E(W_{\underline t_{n+1}}|\underline W_{\underline t_n } )$.
%Usual approaches involve either
% assuming that the series $\{W_{\underline t}\}$ is Markov of order $p$ as in 
%\tr {\cite{pan2016bootstrap}},
%or   approximating 
%$E(W_{\underline t_{n+1}}|\underline W_{\underline t_n } )$
%by a linear function of $\underline{W}_n$ as in 
%\tr {\cite{mcmurry2015high}},
%i.e., contend ourselves with the best linear predictor of $W_{\underline t_{n+1}}$
%denoted by  $\bar E(W_{\underline t_{n+1}}|\underline{W}_{\underline t_n} )$.
%Taking the latter approach,  
%the $L_2$--optimal linear predictor of $W_{n+1}$  based on $\underline{W}_n $ is 
%\begin{equation} \label{NSTS.eq.opt}
% \bar E(W_{n+1}|\underline{W}_n ) = \phi_{1}(n) W_n + \phi_{2}(n) W_{n-1} + \ldots + \phi_{n}(n) W_1,
% \end{equation}
%where the optimal coefficients $\phi_{i}(n)$ are computed from the normal 
%equations, i.e.,  
%$ \phi(n) \equiv (\phi_{1}(n) , \cdots, \phi_{n}(n))'= \Gamma_n^{-1}  \gamma(n)$; 
%here, $\Gamma_n = [\gamma_{|i-j|}]_{i,j=1}^n$ is the autocovariance matrix of
%the  random vector $\underline{W}_n $, and $ \gamma(n) = (\gamma_1, \ldots, \gamma_n)'$  where $\gamma_k=EY_j Y_{j+k} $. 
%Of course,  $\Gamma_n$ is unknown but can be estimated by any of the positive
%definite estimators developed in 
%\tr {\cite{mcmurry2015high}}.
\tb{For this purpose}, the $L_2$--optimal linear predictor of $W_{\underline t_{n+1}}$ can be obtained by fitting
a (causal) AR($p, q$) model to the data $W_{\underline t_1}, \ldots, W_{\underline t_n}$ with $p, q$ chosen by minimizing AIC, BIC or a related criterion as described in \cite{choi2007modeling}; this would entail fitting the model:
\begin{equation} \label{RF.eq.AR}
 % W_t  = \phi_{1}  W_{t-1} + \phi_{2}  W_{t-2} + \cdots + \phi_{p}  W_{t-p}+ V_t
 W_{t_{n_{1}}, t_{n_{2}}} = \sum \limits_{(j,k) \in H_p} \beta_{j,k} W_{t_{n_{1}}-j, t_{n_{2}}-k} + v_{t_1, t_2}
 \end{equation}
where $v_{t_1, t_2}$ is a 2-D white noise process i.e., an uncorrelated sequence, with mean $0$ and variance $\sigma^2 > 0$
% where $V_t$ is a  stationary white noise, i.e., an uncorrelated sequence, with mean zero and variance~$\tau^2$ 
and $(t_{n_{1}}, t_{n_{2}})$ denote the components of $\underline t_{n+1}$.
The   implication then is that 
\begin{equation} 
% \bar E(W_{n+1}|\underline{W}_n ) = \phi_{1}  W_n + \phi_{2}  W_{n-1} + \cdots + \phi_{p}  W_{n-p+1}.
\bar E(W_{\underline t_{n+1}}|\underline{W}_{\underline t_n} ) = \sum \limits_{(j,k) \in H_p} \beta_{j,k} W_{t_{n_{1}}-j, t_{n_{2}}-k}
\label{RF.eq.ARpredictor}
\end{equation} 
%
%As discussed in 
%\tr {the rejoinder to \cite{mcmurry2015high}},
%% McMurry and Politis (2015)
%the two methods for constructing $\bar E(W_{n+1}|\underline{W}_n ) $ are closely related; in fact,   predictor \eqref{NSTS.eq.opt}   coincides with the above AR--type predictor if the matrix    $\Gamma_n$   is the one implied by the
%fitted AR($p$) model \eqref{NSTS.eq.AR}.
% We will use the AR ($p,q$)--type predictor in the sequel because it additionally
% affords us the possibility of resampling based on model~\eqref{RF.eq.AR}.

\subsection{Trend estimation and practical prediction}
\label{RF.sec.trend}

To construct the $L_2$--optimal predictor \eqref{RF.eq.point pred}, 
we need to estimate  the smooth trend $\mu(\cdot)$ and
variance   $\sigma ( \cdot)$
in a nonparametric fashion; this can be easily accomplished via kernel smoothing \tb{by using 2D kernels} ---see 
  e.g. 
% H\"ardle  and Vieu  (1992),   Kim and Cox (1996), or  Li and Racine (2007).
\tr {\cite{hardle1992kernel}, \cite{kim1996bandwidth}, \cite{li2007nonparametric}}.
%When confidence intervals for $\mu( t)$ and  $\sigma (  t)$ 
%are required, however,  matters are more complicated as the asymptotic distribution
%of the different estimators depends on many unknown parameters; see e.g. 
%%  Masry and  Tj{\o}stheim  (1995).
% \tr {\cite{masry1995nonparametric}}.
%Even more difficult is the construction of prediction intervals.
Note, furthermore, that the problem of prediction of $Y_{\underline t_{n+1}}$ involves
estimating the functions \tb{$\mu( \cdot)$ and  $\sigma (\cdot)$}
%$  \mu_{_{[0,1]}}(a,b)$ and  $  \sigma_{_{[0,1]}} (a,b)$  
%described in Remark \ref{RF.re.smooth} for $a=1, b=1$, i.e., it  
is essentially a boundary problem. In such cases, it is well-known that
  local linear   fitting has better properties---in particular, smaller bias---than
kernel smoothing which is well-known to be tantamount to local constant fitting;
% see Fan and Gijbels  (1996), Fan  and  Yao  (2003), or Li and Racine~(2007)
\tr{\cite{fan1996local},\cite{fan2007nonlinear}, or \cite{li2007nonparametric}}.
Note that for time series problems \{$Y_t, \ t \in Z$\} local linear nonparametric estimation can approximate
the trend locally by a straight line whereas for the case of random fields $\{Y_{\underline t}, \ \underline t \in Z^2\}$
discussed in this paper local linear estimation can be used to approximate 
the trend locally with a plane.

\begin{Remark} [One-sided estimation] \rm
Since the goal is predictive inference on $Y_{\underline t_{n+1}}$,
local constant   and/or  local linear fitting must be performed in 
a {\it one-sided way}. 
% To see why, recall that in predictor \eqref{RF.eq.point pred}, the estimands involve $  \mu_{_{[0,1]}}(1)$ and  $ \sigma_{_{[0,1]}} ( 1)$ as just mentioned. 
\tb{Furthermore} to compute $\bar E(W_{\underline t_{n+1}}|\underline{W}_{\underline t_n} )$
in eq.~\eqref{RF.eq.ARpredictor} we need access to the stationary data $W_{\underline t_1},\ldots,W_{\underline t_n}$.
%in order to estimate $\Gamma_n$. 
The $W_{\underline t}$'s are not directly observed, but---much like residuals in a
regression---they can be reconstructed by eq.~\eqref{RF.eq.W} 
with estimates of $\mu(\underline t)$ and   $\sigma (\underline t)$ plugged-in. 
What is important is that   {\bf the way $W_{\underline t}$ is reconstructed/estimated  
by (say) $\hat W_{\underline t}$ must remain the  same for all $\underline t$}, otherwise 
the reconstructed data $\hat W_{\underline t_1},\ldots,\hat W_{\underline t_n}$ can not be considered stationary.
Since $W_{\underline t}$ can only be estimated in a one-sided way for $\underline t$ close to $\underline t_n$,
the same one-sided way must also be implemented for  $\underline t$ in the middle of 
the dataset even though in that case two-sided estimation is possible.    
\label{RF.re.onesided}
\end{Remark}
 
\tb{By}  analogy to model-based regression 
\tr {as described in \cite{Politis2013}},
the one-sided Nadaraya-Watson (NW) kernel estimators of $\mu(\underline t)$ and $\sigma (\underline t )$
can be defined in two ways. 
%In what follows, the notation
%$t_k=k$ will be used; this may appear redundant but
%it makes clear that $t_k$ is the $k$th design point in the
%time series regression, and  allows for easy 
%extension   in the case of missing   data. 
Note that the bandwidth parameter $b$
  will be assumed to satisfy 
\begin{equation}
b\to \infty \ \mbox{as} \ n\to \infty \ \mbox{but} \ b/n\to 0,
\label{RF.eq.new bandwidth}
\end{equation}
 i.e., $b$ is analogous to the product $hn$ \tb{where $h$ is the usual bandwidth in nonparametric regression.} %, see e.g.
We will assume throughout that $K(\cdot)$  is a nonnegative, symmetric 2-D Gaussian kernel function for which the diagonal values are set to the bandwidth $b$ and the off-diagonal terms are set to $0$. Random field data is denoted as $Y_{\underline t_{1}}, \ldots, Y_{\underline t_{k}}, \ldots Y_{\underline t_{n}}$.

\begin{enumerate}

\item  {\bf NW--Regular fitting:}   
% Let $\underline t_{k} \in [\underline t_{b+1}, \underline t_n]$, and define 
Let $\underline t_{k} \in [\underline t_1, \underline t_n]$, and define 

\begin{equation}
\hat \mu(\underline t_k) = \sum _{i=1}^{k} \ Y_{\underline t_{i}} \ \hat K\left(\frac{\underline t_k - \underline t_{i}}{b}\right)
\ \ \mbox{and} \ \ 
\hat M(\underline t_k) = \sum _{i=1}^{k } \ Y_{\underline t_{i}}^2 \ \hat K(\frac{\underline t_k - \underline t_{i}}{b})
\label{RF.eq.nw-mu}
\end{equation}
where
\begin{equation}
 \hat \sigma(\underline t_k) = \sqrt { \hat M_{\underline t_k} - \hat \mu(\underline t_k)^2 }
\ \ \mbox{and} \ \ 
\hat K \left( \frac { \underline t_k - \underline t_{i} } {b}\right) = \frac { K(\frac{\underline t_k - \underline t_{i}} {b}) }{\sum _{j=1}^{k} K(\frac {\underline t_k - \underline t_{j}}{b})} .  
\label{RF.eq.nw-sigma}
\end{equation}
 
Using $\hat \mu(\underline t_k)$ and $\hat \sigma(\underline t_k)$ we can now define the
{\it fitted} residuals by 
\begin{equation}
\hat W_{\underline t_k}= \frac{Y_{\underline t_k}- \hat \mu(\underline t_k)}{ \hat \sigma (\underline t_k )}
\ \ \mbox{for} \ \ 
% \underline t_k=\underline t_{b+1},\ldots, \underline t_n.
\underline t_k=\underline t_{1},\ldots, \underline t_n.
\label{RF.eq.hatW}
\end{equation}
 
\item  {\bf NW--Predictive   fitting (delete-1):}  
Let
%\begin{equation}
%\tilde \mu(t) = \sum _{i=1}^{t-1} \ Y_{i} \ \tilde K\left(\frac{t-t_{i}}{b}\right)
%\ \ \mbox{and} \ \ 
%\tilde M(t) = \sum _{i=1}^{t-1} \ Y_{i}^2 \ \tilde K(\frac{t-t_{i}}{b})
%\label{NSTS.eq.nw-muPRED}
%\end{equation}
%where
%\begin{equation}
% \tilde \sigma(t) = \sqrt { \tilde M_{t} - \tilde \mu(t)^2 }
%\ \ \mbox{and} \ \ 
%\tilde K \left( \frac { t-t_{i} } {b}\right) = \frac { K(\frac{t-t_{i}} {b}) }{\sum _{k=1}^{t-1} K(\frac {t-t_{k}}{b})}   .
%\label{NSTS.eq.nw-sigmaPRED}
%\end{equation}
%%and $K(\cdot)$ is some chosen kernel function. 
%Using $\tilde \mu(t)$ and $\tilde \sigma(t)$ we  now define the
%{\it predictive} residuals by 
%\begin{equation}
%\tilde W_t= \frac{Y_t- \tilde \mu(t)}{ \tilde \sigma (t )}
%\ \ \mbox{for} \ \ 
%t=b+1,\ldots,n.
%\label{NSTS.eq.tildeW}
%\end{equation}

\begin{equation}
\tilde \mu(\underline t_k) = \sum _{i=1}^{k-1} \ Y_{\underline t_{i}} \ \hat K\left(\frac{\underline t_k - \underline t_{i}}{b}\right)
\ \ \mbox{and} \ \ 
\tilde M(\underline t_k) = \sum _{i=1}^{k-1} \ Y_{\underline t_{i}}^2 \ \hat K(\frac{\underline t_k - \underline t_{i}}{b})
\label{RF.eq.nw-muPRED}
\end{equation}
where
\begin{equation}
 \tilde \sigma(\underline t_k) = \sqrt { \hat M_{\underline t_k} - \hat \mu(\underline t_k)^2 }
\ \ \mbox{and} \ \ 
\tilde K \left( \frac { \underline t_k - \underline t_{i} } {b}\right) = \frac { K(\frac{\underline t_k - \underline t_{i}} {b}) }{\sum _{j=1}^{k-1} K(\frac {\underline t_k - \underline t_{j}}{b})} .  
\label{RF.eq.nw-sigmaPRED}
\end{equation}
 
Using $\tilde \mu(\underline t_k)$ and $\tilde \sigma(\underline t_k)$ we can now define the
{\it predictive} residuals by 
\begin{equation}
\tilde W_{\underline t_k}= \frac{Y_{\underline t_k}- \tilde \mu(\underline t_k)}{ \tilde \sigma (\underline t_k )}
\ \ \mbox{for} \ \ 
% \underline t_k=\underline t_{b+1},\ldots, \underline t_n.
\underline t_k=\underline t_{1},\ldots, \underline t_n.
\label{RF.eq.tildeW}
\end{equation}

\end{enumerate}
 \vskip .1in
 \noindent
Similarly, the one-sided local linear (LL) fitting   estimators of $\mu(\underline t_k)$ and $\sigma (\underline t_k )$
can be defined in two ways.

\begin{enumerate}

\item  {\bf LL--Regular fitting:}   
Let $\underline t_{k} \in [\underline t_{1}, \underline t_n]$, and define 
\begin{equation}
\hat \mu(\underline t_k)=\frac{ \sum_{j=1}^{k} w_jY_{\underline t_j} }{\sum_{j=1}^{k } w_j + n^{-2}}
\ \ \mbox{and} \ \ 
\hat M(\underline t_k) = \frac{ \sum_{j=1}^{k} w_jY_{\underline t_j}^2 }{\sum_{j=1}^{k} w_j + n^{-2}}
\label{RF.eq.locallinearF}
\end{equation}
\tb{Denoting} 
\begin{equation}
\underline a = (a_1, a_2) = (\underline t_j - \underline t_k)
\end{equation}

\begin{equation}
s_{t1,1} = \sum_{j=1}^{k} K\left(\frac{\underline t_j - \underline t_{k}}{b}\right) a_1
\end{equation}

\begin{equation}
s_{t2,1} = \sum_{j=1}^{k} K\left(\frac{\underline t_j - \underline t_{k}}{b}\right) a_2
\end{equation}

\begin{equation}
s_{t1,2} = \sum_{j=1}^{k} K\left(\frac{\underline t_j - \underline t_{k}}{b}\right) a_1^2
\end{equation}

\begin{equation}
s_{t2,2} = \sum_{j=1}^{k} K\left(\frac{\underline t_j - \underline t_{k}}{b}\right) a_2^2
\end{equation}

\begin{equation}
s_{t1,t2} = \sum_{j=1}^{k} K\left(\frac{\underline t_j - \underline t_{k}}{b}\right) a_1 a_2
\end{equation}

\begin{equation}
w_j= K(\frac{\underline t_j-\underline t_{k}} {b}) \left[s_{t1,2} s_{t2,2} - s_{t1,t2}^2 - a_1 (s_{t1,1} s_{t2,2} - s_{t2,1} s_{t1,t2}) + a_2 (s_{t1,1} s_{t1,t2} - s_{t1,2} s_{t2,1})
\right],
\label{RF.eq.locallinearweightsF}
\end{equation}
%and
%$ s_{t,k} =\sum_{j=1}^{t }K(\frac{t-t_{j}} {b}) (t-t_{j})^k$
%  for $k=0,1,2$.
 \\
The term $ n^{-2}$  in eq.~\eqref{RF.eq.locallinearF} is just 
to ensure   the denominator is not zero; see    Fan  (1993).
Eq.~\eqref{RF.eq.nw-sigma} then  yields $\hat \sigma(\underline t_k)$,
and eq.~\eqref{RF.eq.hatW} yields~$\hat W_{\underline t_k}$.

\item  {\bf LL--Predictive   fitting (delete-1):}  
Let
 \begin{equation}
\tilde \mu(\underline t_k)=\frac{ \sum_{j=1}^{k-1} w_jY_{\underline t_j} }{\sum_{j=1}^{k-1 } w_j + n^{-2}}
\ \ \mbox{and} \ \ 
\tilde  M(\underline t_k) =\frac{ \sum_{j=1}^{k-1 } w_jY_{\underline t_j}^2 }{\sum_{j=1}^{k-1 } w_j + n^{-2}}
\label{RF.eq.locallinearP}
\end{equation}
where 
\begin{equation}
\underline a = (a_1, a_2) = (\underline t_j - \underline t_k)
\end{equation}

\begin{equation}
s_{t1,1} = \sum_{j=1}^{k-1} K\left(\frac{\underline t_j - \underline t_{k}}{b}\right) a_1
\end{equation}

\begin{equation}
s_{t2,1} = \sum_{j=1}^{k-1} K\left(\frac{\underline t_j - \underline t_{k}}{b}\right) a_2
\end{equation}

\begin{equation}
s_{t1,2} = \sum_{j=1}^{k-1} K\left(\frac{\underline t_j - \underline t_{k}}{b}\right) a_1^2
\end{equation}

\begin{equation}
s_{t2,2} = \sum_{j=1}^{k-1} K\left(\frac{\underline t_j - \underline t_{k}}{b}\right) a_2^2
\end{equation}

\begin{equation}
s_{t1,t2} = \sum_{j=1}^{k-1} K\left(\frac{\underline t_j - \underline t_{k}}{b}\right) a_1 a_2
\end{equation}

\begin{equation}
w_j= K(\frac{\underline t_j-\underline t_{k}} {b}) \left[s_{t1,2} s_{t2,2} - s_{t1,t2}^2 - a_1 (s_{t1,1} s_{t2,2} - s_{t2,1} s_{t1,t2}) + a_2 (s_{t1,1} s_{t1,t2} - s_{t1,2} s_{t2,1})
\right],
\label{RF.eq.locallinearweightsP}
\end{equation}

Eq.~\eqref{RF.eq.nw-sigmaPRED} then  yields $\tilde \sigma(\underline t_k)$,
and eq.~\eqref{RF.eq.tildeW} yields $\tilde W_{\underline t_k}$.

\end{enumerate}

\vskip .1in

\noindent
Using one of the above four methods (NW vs.~LL, regular vs.~predictive)
gives estimates of the quantities needed to compute the $L_2$--optimal 
predictor \eqref{RF.eq.point pred}. In order to 
approximate $E(W_{\underline t_{n+1}}|\underline{Y}_{\underline t_n})$, one would treat the proxies  
$\hat W_{\underline t_k}$ or $\tilde W_{\underline t_k}$ as if they were the true $W_{\underline t_k}$, and proceed
as outlined in Section \ref{RF.sec.TOPP}.
\tb{The bandwidth $b$ in all 4 algorithms described above can be determined by cross-validation as described in Section \ref{RF.cross-validation}.}

\section{Model-free inference} 
\label{RF.Model-free inference}

Model (\ref{RF.eq.model hetero}) is a flexible way to account for a
spatially-changing  mean and variance of $Y_{\underline t}$. 
However, nothing precludes that the random field $\{Y_{\underline t} $ for $\underline t\in {\bf Z^2}   \}$ has a nonstationarity
in its third (or higher moment), and/or in some other feature of its $m$th
marginal distribution.
A way to address this difficulty, and at the same time give a fresh perspective to the problem, is provided by the 
Model-Free Prediction Principle of  Politis (2013, 2015).

The key towards Model-free inference is to be able to construct an invertible transformation 
$H_n: \underline{Y}_{\underline t_n} \mapsto \underline \epsilon_{n}$
where \tb{$ \underline{Y}_{\underline t_n} = (Y_{\underline t_1}, Y_{\underline t_2}, \ldots, Y_{\underline t_n})$ denotes the random field data under consideration}
and $\underline \epsilon_{n}=(\epsilon_{1}, \ldots, \epsilon_{n} )'$
is  a random vector with i.i.d.~components. In order to do this in our context, 
let some $m\geq1$, and denote by ${\mathcal L}(Y_{\underline t_{k} },Y_{\underline t_{k-1}},\ldots,Y_{\underline t_{k-m+1}}) $ 
 the $m$th marginal of the random field { $Y_{\underline t_k}$ }, i.e. the joint probability law of the vector
$(Y_{\underline t_{k} },Y_{\underline t_{k-1}},\ldots,Y_{\underline t_{k-m+1}})'$. Although we  abandon  model~(\ref{RF.eq.model hetero}) 
in what follows, we still want to employ nonparametric smoothing for estimation;  thus,
we must assume that \\${\mathcal L}(Y_{\underline t_{k} },Y_{\underline t_{k-1}},\ldots,Y_{\underline t_{k-m+1}})$ \tor{changes}  smoothly (and slowly) with $\underline t_k$. \tb{In this case $\{Y_{\underline t_k}, \ \underline t_k \in Z^2\}$ can be defined over a 2-D index-set $D$ and the set $Y_{\underline t_{k} },Y_{\underline t_{k-1}},\ldots,Y_{\underline t_{k-m+1}}$ can be considered to be lexicographically ordered as discussed previously in Section \ref{RF.Causality}.}

%\tb {
%\begin{Remark} [Quantifying smoothness--model-free case] \rm
%As in Remark \ref{RF.re.smooth}, we can formally quantify 
%smoothness by mapping the index set $\{1, \ldots, n_1\}$ X $\{1, \ldots, n_2\} $ onto the interval $[0,1]$ X $[0,1]$. 
%Let $\underline{s}=(s_0,s_1,\ldots, s_{m-1})'$, and
% define the distribution function of the $m$th
%marginal by 
% $$D_{\underline t_k }^{(m)}(\underline{s})=P\{ Y_{\underline t_k} \leq s_0,
% Y_{\underline t_{k-1}}\leq s_1,\ldots,Y_{\underline t_{k-m+1} }\leq s_{m-1}  \}. $$ 
%Let $a_{t_1}=(t_1-1)/n_1$ and $b_{t_2}=(t_2-1)/n_2$  where $\underline t_k = (t_1, t_2)$  as before, and   assume that we can write 
%\begin{equation}
%D_{\underline t_k }^{(m)}(\underline{s})=  D^{^{[0,1] X [0,1]}}_{a_{t_1}, b_{t_2}}(\underline{s})
%\ \ \mbox{for} \ t_1=1, \ldots, n_1, t_2=1, \ldots, n_2.
% \label{RF1.eq.smooth2}
%\end{equation}
%We can now quantify smoothness by assuming
% that, for each fixed $\underline{s}$, 
%the function $D^{^{[0,1]}}_{\underline x}(\underline{s})$ is continuous and
%  smooth in $\underline x\in [0,1]$ X $[0,1]$, i.e.,   possesses $k$ continuous derivatives.
%As in Remark~\ref{RF.re.smooth}, here as well it seems to be sufficient
%that $D^{^{[0,1] X [0,1]}}_{\underline x}(\underline{s})$ is   continuous in $\underline x$ but only piecewise smooth. 
%\label{RF.re.smooth2} 
%\end{Remark}
%}

A convenient way to ensure both the smoothness and data-based consistent estimation of
% ${\cal L}(Y_{t },Y_{t-1 },\ldots,Y_{t-m+1})$ is to assume that, \tor{for all t,}}
${\mathcal L}(Y_{\underline t_{k} },Y_{\underline t_{k-1}},\ldots,Y_{\underline t_{k-m+1}})$ is to assume that, for all $\underline t_k,$

\begin{equation}
 Y_{\underline t_k } =
% \overset{\mathrm{\cal L}}{=} 
{\bf f}_{\underline t_k}(W_{\underline t_k },W_{\underline t_{k-1}},\ldots,W_{\underline t_{k-m+1}})
\label{RF.eq.book.9.24}
\end{equation}

\noindent
for some function ${\bf f}_{\underline t_k}$($w$) that is smooth in both arguments $\underline t_k$ and $w$, and some strictly stationary
and weakly dependent,  univariate series ${W_{\underline t_k}}$; without loss of
generality, we may assume that $W_{\underline t_k}$ is a Gaussian series. 
In fact, % Eq. (\ref{RF.eq.book.9.24}) 
Eq. (\ref{RF.eq.book.9.24}) with ${{\bf f}_{\underline t_k}}$($\cdot$) not depending on
$\underline t_k$ is a familiar assumption in studying non-Gaussian and/or long-range dependent 
stationary processes---see e.g. \cite{samorodnitsky1994stable}. 
By allowing ${{\bf f}_{\underline t_k}}$($\cdot$)  to vary smoothly (and slowly) with $\underline t_k$,
% Eq. (\ref{RF.eq.book.9.24})  
Eq. (\ref{RF.eq.book.9.24}) can be used to describe a rather general
class of locally stationary processes.
% In the above the symbol
%$\overset{\mathrm{\mathcal L}}{=}$ denotes equality in distribution, i.e. the left-hand side of Eq. (\ref{NSTS1.eq.book.9.24})
%has the same probability law as the right-hand side. 
Note that model~(\ref{RF.eq.model hetero}) is a special case
of Eq. (\ref{RF.eq.book.9.24}) with $m=1$, and
 the function ${{\bf f}_{\underline t_k}}$($w$) being affine/linear in $w$.
% and $\overset{\mathrm{\mathcal L}}{=}$ being replaced by usual equality. 
Thus, for concreteness and easy comparison with the model-based case of
Eq. (\ref{RF.eq.model hetero}), we will focus in the sequel on the case $m=1$. 
For reference model-free estimators for point prediction and prediction intervals in the case of locally stationary time series for $m=1$ have been discussed in \cite{das2021predictive}.
% Section \ref{NSTS.sec.higher-dimensional marginals} discusses how to handle the case $m > 1$.

%

\subsection{Constructing the theoretical transformation}
\label{RF.sec.CTT}

\tb {Hereafter, adopt the setup of \tor{ Eq. (\ref{RF.eq.book.9.24})} with $m=1$,}
and let 
$$D_{\underline t}(y)=P\{ Y_{\underline t} \leq y  \}$$ denote  the 1st marginal distribution of random field $\{Y_{\underline t} \}$.
Throughout Section \ref{RF.Model-free inference},
the default assumption will be that $D_{\underline t }(y)$ is (absolutely) continuous in $y$ for all $\underline t$. %however, a departure from this assumption will be discussed  in Section \ref{NSTS.discrete data}.  

We now define new variables via the probability integral transform, i.e., let 
\begin{equation}
U_{\underline t} =  D_{\underline t}(Y_{\underline t}) \ \ \mbox{for} \ \underline t= \underline t_1,\ldots, \underline t_n;
 \label{RF_unif.eq.modelT}
\end{equation}
the assumed continuity of $D_{\underline t }(y)$  in $y$  implies that 
  $U_{\underline t_1},\ldots, U_{\underline t_n}$ are random variables having   distribution Uniform $ (0,1)$.
However, $U_{\underline t_1},\ldots, U_{\underline t_n}$ are dependent; to transform them to 
independence, a preliminary transformation towards Gaussianity is helpful as
 discussed in 
\tr{\cite{Politis2013}}.
Letting $\Phi$ denote the cumulative distribution function (cdf) of the standard normal distribution, we define
\begin{equation}
Z_{\underline t} = \Phi^{-1} (U_{\underline t})  \ \ \mbox{for} \ \underline t=\underline t_1,\ldots, \underline t_n;
 \label{RF_norm.eq.modelT}
\end{equation}
it then follows that $Z_{\underline t_1},\ldots, Z_{\underline t_n}$ are standard normal---albeit 
correlated---random variables. 

Let $ \Gamma_n $ denote  the $n\times n$ covariance matrix
of the   random vector $\underline{Z}_{\underline t_n}=(Z_{\underline t_1},\ldots, Z_{\underline t_n})' $.
Under standard   assumptions,
e.g. that the spectral density of the series $\{Z_{\underline t_n}\}$
is continuous and bounded away from zero,\footnote{If the spectral density 
is equal to zero over an interval---however small---then the time series $\{Z_{\underline t_n}\}$ is perfectly predictable 
based on its infinite past, and the same would be true for 
 the time series $\{Y_{\underline t_n}\}$; see Brockwell and Davis (1991, Theorem 5.8.1)
on Kolmogorov's formula.}   the matrix $ \Gamma_n $ is invertible 
when  $n$ is large enough. Consider the 
Cholesky decomposition   $ \Gamma_n  =  C_n   C_n'$ 
where $  C_n$ is (lower) triangular, and  construct the 
{\it whitening} transformation:
\begin{equation}
\label{RF.eq.whitenfilterT}
 \underline \epsilon_{n}=   C_n^{-1} \underline{Z}_{\underline t_n} .
% \ \ \mbox{and hence} \ \ 
% \underline{Z}_n =   C_n \underline \epsilon_{n} .
\end{equation}
It then follows that the entries of $\underline \epsilon_{n}=(\epsilon_1, \ldots, \epsilon_n)'$ are uncorrelated~standard normal.
Assuming that the random variables $Z_{\underline t_1},\ldots, Z_{\underline t_n}$ were {\it jointly} normal, 
this can be strenghtened to claim that $
\epsilon_1, \ldots, \epsilon_n$ are i.i.d.~$N(0,1)$.
\tb{Joint normality can be established by assuming a generative model of the random field as given by eq.(\ref{RF.eq.book.9.24}),
for a more detailed discussion refer to \cite{das2021predictive}}.
% see  Section \ref{NSTS.sec.higher-dimensional marginals} for further discussion.
Consequently, the transformation 
of the dataset % $\underline{Y}_n=(Y_1,\ldots, Y_n)' $
\tb{$ \underline{Y}_{\underline t_n} = (Y_{\underline t_1}, Y_{\underline t_2}, \ldots, Y_{\underline t_n})$} 
to the vector $\underline \epsilon_{n}$ with i.i.d.~components has been
achieved as required in premise (a) of the 
Model-free Prediction Principle. % of Politis (2013) 
 Note that all the steps in the transformation, i.e., 
   eqs.~(\ref{RF_unif.eq.modelT}), (\ref{RF_norm.eq.modelT}) and (\ref{RF.eq.whitenfilterT}), are invertible; hence, the composite
transformation % $H_n: \underline{Y}_n \mapsto \underline \epsilon_{n}$
\tb{$H_n: \underline{Y}_{\underline t_n} \mapsto \underline \epsilon_{n}$}
is invertible as well.

\subsection{Kernel estimation of the `uniformizing' transformation}
\label{RF.sec.KEUT}

We first focus on estimating the `uniformizing' part of the transformation,
i.e., eq.~(\ref{RF_unif.eq.modelT}).
Recall that the Model-free setup implies that the function $D_{\underline t}(\cdot)$ changes smoothly (and slowly) with~$\underline t$; hence, local constant   and/or local linear fitting can be used to estimate it. Consider random field data denoted as $Y_{\underline t_{1}}, \ldots, Y_{\underline t_{k}}, \ldots Y_{\underline t_{n}}$.
Using local constant, i.e.,  kernel estimation, a consistent  estimator of the marginal distribution 
$D_{\underline t_k }(y)$ is given by:
\begin{equation}
\hat D_{\underline t_k }(y) = \sum_{i=1}^{T} {\bf 1}\{ Y_{\underline t_{i}}\leq y\} 
 \tilde K (\frac{\underline t_{k} - \underline t_{i}}{b})
\label{RF.eq.hatD}
\end{equation}
where $\tilde K (\frac{\underline t_k -\underline t_{i}}{b}) =  K (\frac{\underline t_k -\underline t_{i}}{b})/
  \sum_{j=1}^{T}K (\frac{\underline t_k-\underline t_{j}}{b}) $. Similar to the model-based case we will assume throughout that $K(\cdot)$  is a nonnegative, symmetric 2-D Gaussian kernel function for which the diagonal values are set to the bandwidth $b$ and the off-diagonal terms are set to $0$.
Note that the kernel estimator \eqref{RF.eq.hatD} is {\it one-sided}
for the same reasons discussed in Remark \ref{RF.re.onesided}.
Since $\hat D_{\underline t_k }(y)$ is a step function in $y$, a smooth estimator 
can be defined as:
\begin{equation}
\bar D_{\underline t_k }(y) = \sum_{i=1}^{T} \Lambda(\frac {y-Y_{\underline t_{i}}} {{h}_0})\tilde K (\frac{\underline t_k-{\underline t_i}}{b})
\label{RF.eq.barD}
\end{equation}
where ${h}_0$ is a secondary bandwidth.
 Furthermore, as in Section \ref{RF.sec.trend}, we can let $T=k$ or $T=k-1$
leading to a {\bf fitted vs.~predictive} way to estimate $D_{\underline t_k }(y)$
by either $\hat D_{\underline t_k }(y)$  or $\bar D_{\underline t_k }(y)$. 
\tr {Cross-validation is used to determine the bandwidths $h_0$ and $b$ \tor{; details} 
are described in Section \ref{RF.cross-validation}.}

% \begin{Remark} [On   choice of the two bandwidths] \rm
% To choose the main bandwidth $b$ for either
%$\hat D_{t }(y)$  or $\bar D_{t }(y)$, predictive cross-validation   may be used
%but  it must be adapted to the time series prediction setting, i.e.,
%as in Remark \ref{NSTS.re.bandwidthCV}. 
%Define   $h=b/n$, and recall that in the analogous
%regression problem 
% % see Li and Racine (2007)
% \tr {\cite {Politis2013}}
%the optimal rates $h_0\sim n^{-2/5}$ and $h \sim n^{-1/5}$ 
%were suggested in connection with the nonnegative kernel $K$;
%this led to the practical recommendation of letting $h_0=h^2$
%with $h$ chosen by cross-validation. Similarly here, 
%the recommendation is to choose 
%  $b$     by   the time series
%cross-validation of Remark \ref{NSTS.re.bandwidthCV}, 
%and then let ${h}_0=b^2/n^2$. 
%\label{NSTS.re.bandwidthCV2}
%\end{Remark}

%%%%%%%%%%%%%%%%%%%%%%%%%%%%%%%%%%%%%%%%%%%%%%%%%%%%%%%%%%%%%%%%%%%%%%%%%%
\subsection{Local linear estimation of the `uniformizing' transformation}
\label{RF.sec.LLEUT}

Note that the kernel estimator 
$\hat D_{\underline t_k }(y)$ defined in eq.~\eqref{RF.eq.hatD}
is just the Nadaraya-Watson smoother, i.e., local average,  of the variables 
$u_1,\ldots, u_n$ where $u_i = {\bf 1}\{ Y_{\underline t_i}\leq y\}$.
 Similarly, $\bar D_{\underline t _k}(y)$ defined in eq.~\eqref{RF.eq.barD}
is just the Nadaraya-Watson smoother of the variables 
$v_1,\ldots, v_n$ where $v_i =\Lambda(\frac {y-Y_{\underline t_i}} {{h}_0})$. 
In either case, it is only natural to try to consider a local
linear smoother as an alternative to Nadaraya-Watson especially
\tb{since, once again, our interest lies in one-sided estimation on the boundary of the random field.}
% on the boundary, i.e., the case $t=n$. 
  
Let 
\tr {$\hat D_{\underline t_k }^{LL}(y)$ and $\bar D_{\underline t_k}^{LL}(y)$} 
denote the local linear  estimators of $D_{\underline t_k }(y)$
based on either the indicator variables ${\bf 1}\{ Y_{\underline t_i}\leq y\}$
or the smoothed variables $\Lambda(\frac {y-Y_{\underline t_i}} {{h}_0})$ 
\tr {respectively}. 
Keeping $y$ fixed, 
\tr {$\hat D_{\underline t_k }^{LL}(y)$ and $\bar D_{\underline t_k}^{LL}(y)$}
\tr {exhibit} good behavior \tr{for estimation at the boundary,} e.g.
smaller bias than either $\hat D_{\underline t_k }(y)$ \tr{and} $\bar D_{\underline t_k }(y)$
\tr{respectively}.
However, there is no guarantee that
\tr {these will be}
proper distribution \tr{functions} as a function of $y$, 
i.e., being nondecreasing in $y$ with a left limit of 0 and
a right limit of 1; see 
\tr{\cite{li2007nonparametric}}
for a discussion.

%There have been several proposals in the literature to address this issue.
%An interesting one   is the adjusted Nadaraya-Watson estimator 
%of 
%% Hall et al.~(1999) 
%\tr {\cite{Hall1999}}
%which, however, is   tailored
%towards nonparametric autoregression estimation rather than our 
%setting where $Y_t$ is regressed on $t$. 
%Coupled with the fact that we are interested in the 
%boundary case $t=n$, the equation yielding the adjusted Nadaraya-Watson  
%  weights   \tb{do not always} admit a solution. 

\tr {One proposed solution put forward by
\cite{hansen2004nonparametric}} 
 \tr {involves}
 a straightforward adjustment to 
the local linear estimator of a conditional distribution function 
that maintains its favorable asymptotic properties.
The local linear  versions of $\hat D_{\underline t_k}(y) $ and $\bar D_{\underline t_k}(y) $
adjusted via 
  Hansen's (2004) 
%\tr{\cite{hansen2004nonparametric}}
proposal are
  given as follows:
\begin{equation}
\label{RF.eq.sll_cdf}
\hat D_{\underline t_k}^{LLH}(y) = \frac{\sum_{i=1}^{T} w_{i}^\diamond {\bf 1}(Y_{{\underline t_i}} \le y)}{\sum_{i=1}^{T} w_{i}^\diamond}  
\ \ \mbox{and} \ \ 
\bar D_{\underline t_k}^{LLH}(y) = \frac{\sum_{i=1}^{T} w_{i}^\diamond 
\Lambda(\frac {y-Y_{\underline t_i}} {{h}_0})}{\sum_{i=1}^{T} w_{i}^\diamond} .
\end{equation}
%The weights $w_{i}^\diamond $ are defined by
%\begin{align}
%\label{NSTS.eq.sll_diamond}
%w_{i}^\diamond = \begin{cases}
%                 \ 0  & \ \mbox{when} \ \hat \beta(t-t_{i}) > 1 \\
%                 \ w_{i}(1 - \hat \beta(t-t_{i})) & \ \mbox{when} \ \hat \beta(t-t_{i}) \le 1
%                 \end{cases}
%\end{align}
%where 
%\begin{equation}
%\label{NSTS.eq.sll_weights}
%w_{i} = \frac{1}{b}\ K(\frac{t-t_{i}}{b})
%\ \ \mbox{and} \ \ 
%\hat \beta =\frac{\sum_{i=1}^T\ w_{i}(t-t_{i})}{ \sum_{i=1}^T\ w_{i}(t-t_{i})^2 }.
%\end{equation}

The weights $w_{i}^\diamond$ are derived from weights $w_i$ described in equations (\ref{RF.eq.locallinearweightsF}) and (\ref{RF.eq.locallinearweightsP}) for the fitted and predictive cases where:
\begin{align}
\label{RF.eq.sll_diamond}
w_{i}^\diamond = \begin{cases}
                 \ 0  & \ \mbox{when} \ \ w_i < 0 \\
                \ w_{i} & \ \mbox{when} \ \ w_i \ge 0
                 \end{cases}
\end{align}

\tr 
{As with eq.~\eqref{RF.eq.hatD}and~\eqref{RF.eq.barD}, we can let $T=k$ or $T=k-1$ in the above, 
leading to a {\bf fitted vs.~predictive} 
local linear estimators of  $D_{\underline t_k }(y)$,
by either $\hat D_{\underline t_k}^{LLH}(y)$ or $\bar D_{\underline t_k}^{LLH}(y)$}.

\tr{
\subsection{Uniformization using Monotone Local Linear Distribution Estimation}
}
\label{RF.sec.MLLDE}

\tr
{Hansen's \tb {(2004)} proposal replaces  negative weights by zeros, and then renormalizes the nonzero
weights. The problem here is that if estimation is performed on the boundary (as in the case
with \tor{one-step ahead prediction of} random fields), negative weights are crucially needed in order to ensure \tor{the}
extrapolation takes place with minimal bias.} 
\tc{A recent proposal by  \cite{das2019nonparametric}
addresses this issue by modifying the original, possibly nonmonotonic local linear distribution estimator $\bar D_{\underline t_k}^{LL }(y) $ to construct a monotonic version 
denoted by $\bar {D}_{\underline t_k}^{LLM }(y)$.}

\newpage
\tc{The Monotone Local Linear Distribution Estimator $\bar {D}_{\underline t_k}^{LLM }(y)$ can be   constructed by Algorithm \ref{Monotone_Density_Algo} given below.}

\medskip

%\begin{Algorithm} {
%\bf{Monotone Local Linear Distribution Estimation - Method 1}
%\label{Monotone_Algo}
%}
%\label{Monotone_LL}
%\begin{enumerate} 
%\item Compute  $\bar D_{t}^{LL }(y) $,  and denote $l=\lim_{y\to - \infty} \bar {D}_{t}^{LL }(y) $. 
%\item Define a function $G_1(y)=  \bar {D}_{t}^{LL }(y)-l$. 
%\item  Define a second function $G_2$ with the property that $G_2(y+\epsilon)=  \max 
%\left( G_1(y+\epsilon), G_1(y) \right)$ for all $y$ and all $\epsilon >0$.
%\item Define $\bar {D}_{t}^{LLM }(y) =G_2(y)/L$ where   $L=\lim_{y\to   \infty} G_2(y) $. 
%\end{enumerate}
%\end{Algorithm}
%
%
%\tc{A different---albeit equivalent---way of constructing the estimator
% $\bar {D}_{t}^{LLM }(y) $ is
%by first constructing its associated density function denoted by $\bar d_{t}^{LLM }(y) $
%which  will be called the {\it Monotone Local Linear
%Density Estimator}. 
%The alternative algorithm goes as follows.}

\begin{Algorithm}{
\bf{Monotone Local Linear Distribution Estimation}
\label{Monotone_Density_Algo}
}
\label{Monotone_density_LL}
\begin{enumerate}
\item Recall that the derivative of $\bar {D}_{\underline t_k}^{LL  }(y) $ 
with respect to $y$ is given by 
$$  \bar d_{\underline t_k}^{LL}(y)=\frac{\frac {1} {{h}_0} \sum_{j=1}^{T } w_j  \lambda(\frac {y-Y_{ {\underline t_j}}} {{h}_0})  }{\sum_{j=1}^{T} w_j  }  
 $$
where $\lambda(y)$ is the derivative of  $\Lambda(y)$ and the weights $w_j$ can be derived based on equations (\ref{RF.eq.locallinearweightsF}) and (\ref{RF.eq.locallinearweightsP}) for the fitted and predictive cases.
\item Define a nonnegative version of $  \bar d_{\underline t_k}^{LL}(y)$ as
$  \bar d_{\underline t_k}^{LL+}(y)=\max (\bar d_{\underline t_k}^{LL}(y), 0)$.
\item To make the above a proper density function, renormalize
it to area one, i.e., let 
\begin{equation}
\label{eq.densityMLL}
\bar d_{\underline t_k}^{LLM }(y) = \frac{\bar d_{\underline t_k}^{LL+}(y)  } {\int_{-\infty}^\infty \bar d_{\underline t_k}^{LL+}(s)ds  }.
\end{equation}
\item Finally, define $\bar {D}_{\underline t_k}^{LLM }(y) =\int_{-\infty}^y \bar d_{\underline t_k}^{LLM }(s) ds.$
\end{enumerate}
\end{Algorithm}

 The above modification of the local linear estimator
allows one to maintain monotonicity while retaining the negative weights that
are helpful in problems which involve estimation at the boundary.
{As with eq.~\eqref{RF.eq.hatD} and~\eqref{RF.eq.barD}, we can let $T=k$ or $T=k-1$ in the above, 
leading to a {\bf fitted vs.~predictive} 
local linear estimators of  $D_{\underline t_k }(y)$ that are monotone.
}

Different algorithms could also be employed for performing monotonicity correction on the original estimator $\bar {D}_{\underline t_k}^{LL  }(y) $; these are discussed in detail in  \cite{das2019nonparametric}.  In practice, Algorithm \ref{Monotone_Density_Algo} is
preferable because it is the fastest in term of implementation; notably, 
 density estimates can be obtained in a fast way (using the Fast Fourier Transform) 
using standard functions in statistical software such as R. 
% Computational speed is particularly important 
% in constructing
% bootstrap prediction intervals
Computational speed is  important 
\tb{in point prediction but is critical for cross-validation
 where a large number of estimates of $\bar D_{\underline t_k }^{LLM}(y)$ must be computed to determine the optimal bandwidth.}

\subsection{Estimation of the whitening transformation} 
\label{RF.seq.whiteningtransformation}

To implement the whitening transformation \eqref{RF.eq.whitenfilterT},
it is necessary to estimate $ \Gamma_n $, i.e., the $n\times n$ covariance matrix
of the   random vector $\underline{Z}_{\underline t_n}=(Z_{\underline t_1},\ldots, Z_{\underline t_n})' $
where the $Z_{\underline t}$ are the normal random variables
 defined in eq.~\eqref{RF_norm.eq.modelT}. 
% Let $\{Z_{\underline t} = Z_{\underline t_1, \underline t_2} | t_1 = 1, \ldots, T_1,  t_2 = 1, \ldots, T_2,  n=T_1T_2\}$

%As discussed in the analogous \tr {model-based} problem in Section \ref{NSTS.sec.TOPP}, there
%are two approaches towards positive definite estimation of $ \Gamma_n $ based on the sample  $Z_1,\ldots, Z_n$. They are both based on the sample autocovariance 
%defined as $\breve \gamma_k = n^{-1}\sum_{t=1}^{n-|k|}Z_tZ_{t+|k|}$
%for $|k|<n$; for $|k|\geq n$, we define $\breve \gamma_k =0$.  

%\tb{The problem involves positive definite estimation of $ \Gamma_n $ based on the sample  $Z_{\underline t_1},\ldots, Z_{\underline t_n}$. 
%For this purpose fit a causal AR($p, q$) model to the data $Z_{t_1},\ldots, Z_{t_n}$ with $p, q$
%by minimizing AIC, BIC or a related criterion as described in \cite{choi2007modeling}. Then, let $\hat \Gamma_n^{AR}$ be the 
%$n\times n$ covariance matrix associated with the fitted AR model. Let $\hat \gamma_{|i-j|}^{AR}$ denote the $i,j$  element of the
%Toeplitz matrix $\hat \Gamma_n^{AR}$. Using the 2D Yule-Walker
%equations described in \cite{choi2007modeling} to fit the AR model implies that  $\hat \gamma_{k}^{AR}= \breve \gamma_k $ for $k=0,1,\ldots, p.$ %p-1$.
%For $k> p$, $\hat \gamma_{k}^{AR}$ can be found by solving
%(or just iterating) the 
%difference equation that characterizes the (fitted) AR model;
%R automates this process via the {\tt ARMAacf()} function.}

\tb{The problem involves positive definite estimation of $ \Gamma_n $ based on the sample  $Z_{\underline t_1},\ldots, Z_{\underline t_n}$. 
Let $\hat \Gamma_n^{AR}$ be the  $n\times n$ covariance matrix associated with the fitted AR(p,q) model to the data $Z_{\underline t_1},\ldots, Z_{\underline t_n}$. with $p, q$
by minimizing AIC, BIC or a related criterion as described in \cite{choi2007modeling}. Let $\hat \gamma_{|i-j|}^{AR}$ denote the $i,j$  element of the Toeplitz matrix $\hat \Gamma_n^{AR}$. Using the 2D Yule-Walker equations to fit the AR model implies that $\hat \gamma_{k, l}^{AR}= \breve \gamma_{k,l} $ for $k=0,1,\ldots, p$ and $l=0,1, \ldots, q$. For the cases where $ k > p$ or $l > q, \hat \gamma_{k, l}^{AR}$ can be fitted by iterating the difference equation that characterizes the fitted 2D AR model. In the R software this procedure is automated for time series using the  {\tt ARMAacf()} function, here we extend the same for stationary data over random fields.}

\noindent Estimating the `uniformizing' transformation $D_{\underline t}(\cdot)$ and  the whitening transformation based on $  \Gamma_n$ 
allows us to estimate the  transformation $H_n: \underline{Y}_{\underline t_n} \mapsto \underline \epsilon_{n}$. However, in order to put the Model-Free Prediction Principle to work, we also need to   estimate the transformation $H_{n+1}$ (and   its inverse). \tb{To do so, we need a positive definite estimator for 
the matrix $  \Gamma_{n+1}$; this can be accomplished by extending the covariance matrix associated with the fitted 2D AR(p,q) model to $(n+1) by (n+1)$ i.e. calculate $\hat \Gamma_{n+1}^{AR}$.}

 \noindent
Consider the `augmented' vectors:
\begin{itemize}
\item $\underline{Y}_{\underline t_{n+1}}= (Y_{\underline t_1},\ldots,Y_{\underline t_n}, Y_{\underline t_{n+1}})'$, 
\item $\underline{Z}_{\underline t_{n+1}}=(Z_{\underline t_1}, \ldots,Z_{\underline t_n},   Z_{\underline t_{n+1}})'$
and 
\item $\underline \epsilon_{n+1}=(\epsilon_1,\ldots, \epsilon_n, \epsilon_{n+1})'$
\end{itemize}
where the values $Y_{\underline t_{n+1}}, Z_{\underline t_{n+1}}$ and $\epsilon_{n+1}$
are yet unobserved. 
We now show how to obtain 
the inverse transformation $H_{n+1}^{-1}: 
\underline \epsilon_{n+1} \mapsto 
\underline{Y}_{\underline t_{n+1}} $. Recall that $\underline \epsilon_{n}$ and $
\underline{Y}_{\underline t_n } $ are related in a one-to-one way via 
  transformation $H_{n}$, so the values $ Y_{\underline t_1},\ldots,Y_{\underline t_n}$
are obtainable by $\underline{Y}_{\underline t_n } =H_{n}^{-1}(\epsilon_{n}).$
Hence, we just need to show how to create the unobserved
$Y_{\underline t_{n+1}}$   from $\underline \epsilon_{n+1} $; this is 
done in the following three steps.

\begin{Algorithm} {
\tor{GENERATION OF UNOBSERVED DATA\tor{POINT} FROM \tor{FUTURE }INNOVATIONS}
\label{Pred_MF_Algo}
}

\begin{itemize}
 
 \smallskip
 \item [i.] Let 
\begin{equation}
\label{RF.eq.CholPred}
\underline{Z}_{\underline t_{n+1}}=  C_{n+1} \underline \epsilon_{n+1}
\end{equation}
 where $  C_{n+1}$ is the (lower) triangular Cholesky factor of
(our positive definite estimate of)  $  \Gamma_{n+1} $.
From the above, it follows that 
 
\begin{equation}
\label{RF.eq.invtrans1}
{Z}_{\underline t_{n+1}}=\underline c_{n+1} \underline \epsilon_{n+1}
\end{equation}
where $\underline c_{n+1}=(c_1,\ldots,c_n,c_{n+1}) $ is 
a row vector consisting of  the last row of
matrix $  C_{n+1}$.

\item [ii.] Create the uniform random variable 
\begin{equation}
\label{RF.eq.invtrans2}
{U}_{\underline t_{n+1}}= \Phi(Z_{\underline t_{n+1}}) . 
\end{equation}

\item [iii.] 
Finally, define  
\begin{equation}
\label{RF.eq.invtrans3}
Y_{\underline t_{n+1}} =   D_{n+1}^{-1} ({U}_{\underline t_{n+1}}); 
\end{equation}
of course, in practice, the above will be based on an
estimate of  $ D_{n+1}^{-1}(\cdot)$. 
\end{itemize}
\end{Algorithm}

\noindent
Since $\underline{Y}_{\underline t_n } $  has already been created using (the
first $n$ coordinates of) $\underline \epsilon_{n+1}$, the above completes
the construction of $\underline{Y}_{\underline t_{n+1}} $ based on $\underline \epsilon_{n+1}$,
i.e., the mapping  $H_{n+1}^{-1}: 
\underline \epsilon_{n+1} \mapsto \underline{Y}_{\underline t_{n+1}} $. 
%H_n:  eqs.~(\ref{NSTS1_unif.eq.modelT}), (\ref{NSTS1_norm.eq.modelT}) and (\ref{NSTS.eq.whitenfilterT}),

%%%%%%%%%%%%%%%%%%%%%%%%%%%%%%%%%%%%%%%%%%%%%%%%%%%%%%%%%%%%
% \subsection{Model-free predictors and   prediction intervals}
\subsection{Model-free point prediction}
\label{RF.sec.prediction intervals}

In the previous sections, it was shown how the construct the  transformation 
$H_n: \underline{Y}_{\underline t_n} \mapsto \underline \epsilon_{n}$  and its
inverse $H_{n+1}^{-1}: 
\underline \epsilon_{n+1} \mapsto \underline{Y}_{\underline t_{n+1}} $,
where the random variables $\epsilon_{1},  \epsilon_2,  \ldots,$
are i.i.d. Note that by combining eq.~(\ref{RF.eq.invtrans1}), (\ref{RF.eq.invtrans2}) and (\ref{RF.eq.invtrans3}) we can write the formula: 
$$
  Y_{\underline t_{n+1}} =   D_{n+1}^{-1}\left( \Phi( \ \underline c_{n+1} \underline \epsilon_{n+1} )\right).
$$
Recall that $\underline c_{n+1} \underline \epsilon_{n+1}
= \sum_{i=1}^n c_i \epsilon_i+c_{n+1} \epsilon_{n+1} $;
hence, the above can be compactly denoted as 
\begin{equation}
\label{RF.eq.pred.equation}
  Y_{\underline t_{n+1}} =  g_{n+1}(\epsilon_{n+1}) 
\ \ \mbox{where} \ \ 
g_{n+1}(x)=
 D_{\underline t_{n+1}}^{-1}\left( \Phi \left( \  \sum_{i=1}^n c_i \epsilon_i+c_{n+1}x \right) \right).
\end{equation}
Eq.~\eqref{RF.eq.pred.equation} is the predictive equation required in the 
Model-free Prediction Principle; % of Politis (2013)
conditionally on $\underline{Y}_{\underline t_n } $, it can be used like a model
equation in computing the $L_2$-- and $L_1$--optimal point predictors of 
 $Y_{\underline t_{n+1}} $. We will give these in detail as part of the 
 % general algorithms for the construction of Model-free predictors and prediction intervals.
\tb{general algorithm for the construction of Model-free point predictors.}

\begin{Algorithm} 
\label{RF.Algo.BasicMF}
% {\sc   Model-free (MF) predictors and prediction intervals for $Y_{\underline t_{n+1}} $}
{\sc   Model-free (MF) point predictors for $Y_{\underline t_{n+1}} $}
 
\begin{enumerate}
\item Construct $U_{\underline t_1},\ldots, U_{\underline t_n}$ by eq.~(\ref{RF_unif.eq.modelT}) with 
$D_{\underline t_n}(\cdot)$ estimated by either $\bar D_{\underline t_n}(\cdot)$ \tr {, $\bar D_{\underline t_n}^{LLH}(\cdot)$
or $\bar D_{\underline t_n}^{LLM}(\cdot)$};
for \tr {all the 3 types of estimators}, \tb{use the respective formulas with $T=k$.}
\item Construct $Z_{\underline t_1},\ldots, Z_{\underline t_n}$ by eq.~(\ref{RF_norm.eq.modelT}), and use the
methods of Section \ref{RF.seq.whiteningtransformation} to estimate
$\Gamma_n$ by 
% either $\hat \Gamma_n^{AR}$ or 
% $\hat \Gamma_n^\star$.
\tb{$\hat \Gamma_n^{AR}$.}
\item  
Construct $\epsilon_1,\ldots, \epsilon_n$ by eq.~(\ref{RF.eq.whitenfilterT}),
and let $\hat F_n$ denote their empirical distribution.

\item 
The Model-free $L_2$--optimal point predictor  of 
 $Y_{\underline t_{n+1}} $ is then \\
\tr{$$ \hat Y_{\underline t_{n+1}}= \int  g_{n+1}(x ) dF_n(x)
= \frac{1}{n} \sum_{i=1}^n g_{n+1}(\epsilon_i ) $$}
where the function $g_{n+1}$ is defined in the 
predictive equation \eqref{RF.eq.pred.equation}
with $D_{\underline t_{n+1}}  (\cdot)$ being again estimated by either $\bar D_{\underline t_{n+1}}  (\cdot)$
\tr{
,
$ \bar D_{\underline t_{n+1}}^{LLH}  (\cdot)$
or $ \bar D_{\underline t_{n+1}}^{LLM}  (\cdot)$
}  
\tr {all} 
\tb{with $T=k$.}

\item 
The Model-free $L_1$--optimal point predictor  of 
 $Y_{\underline t_{n+1}} $ is given by the median of the set
$\{  g_{n+1}(\epsilon_i )$ for $i=1,\ldots, n\}$.

%\item 
%Prediction intervals for $Y_{\underline t_{n+1}} $ with prespecified coverage
%probability can be constructed 
%%via the Model-free Boootstrap of   Algorithm \ref{MF3short.Algorithm1} % Algorithm 2.4.1 of Politis (2015)--the book! 
%% This is Algorithm A.1 in the APPENDIX TO THE PAPER 
%based on either the $L_2$-- or $L_1$--optimal point predictor.
\end{enumerate}
\end{Algorithm} 
\vskip .13in
\noindent
Algorithm \ref{RF.Algo.BasicMF}  used the construction of
$\bar D_{\underline t_k}(\cdot)$ \tr {,
$\bar D_{\underline t_k}^{LLH}(\cdot)$ 
\tr {or $\bar D_{\underline t_k}^{LLM}(\cdot)$ }
}
  with $T=k$;  using $T=k-1$ instead,  
 leads to the {\it predictive} version of the algorithm.
\vskip .173in
 \begin{Algorithm} 
\label{RF.Algo.PMF}
% {\sc   Predictive Model-free (PMF) predictors and prediction intervals for $Y_{\underline t_{n+1}} $} \\
{\sc   Predictive Model-free (PMF) predictors for $Y_{\underline t_{n+1}} $} \\
The algorithm is identical to Algorithm \ref{RF.Algo.BasicMF}
except for using $T=k-1$ instead of $T=k $ in the construction
of $\bar D_{\underline t_k}(\cdot)$ 
\tr{
,  $\bar D_{\underline t_k}^{LLH}(\cdot)$ and $\bar D_{\underline t_k}^{LLM}(\cdot)$.
}
 \end{Algorithm} 
\vskip .13in

\section{Random Fields  cross-validation}
\label{RF.cross-validation}

% \begin{Remark} [Random Fields  cross-validation] \rm
 To choose the  bandwidth $b$ for either % of the above methods,
 \tb{model-based or model-free point prediction}
 predictive cross-validation may be used
but  it must be adapted to the random field prediction setting, i.e., always 
one-step-ahead. To elaborate, let  $k<n$, and suppose only 
subseries $Y_{\underline t_1},\ldots, Y_{\underline t_k}$ has been observed.
Denote
$\hat Y_{\underline t_{k+1}}$ the best predictor of $Y_{\underline t_{k+1}}$ based on the
data $Y_{\underline t_1},\ldots, Y_{\underline t_k}$  constructed according to the
above methodology and some choice of $b$. However, since $Y_{\underline t_{k+1}}$ is known, the
quality of the predictor can be assessed. So, for each value of $b$
over a reasonable range,  we can form either
 $PRESS(b)=\sum_{k=k_o}^{n-1} ( \hat Y_{\underline t_{k+1}} - Y_{\underline t_{k+1}})^2 $
or  $PRESAR(b)=\sum_{k=k_o}^{n-1} | \hat Y_{\underline t_{k+1}} - Y_{\underline t_{k+1}}| $; here
$k_o$ should be big enough so that estimation is accurate,
e.g., $k_o$ can be of the order of $\sqrt{n}$.
The cross-validated  bandwidth choice would then be the $b$ that
minimizes $PRESS(b)$; alternatively, we can choose to minimize $PRESAR(b)$ if
 an   $L_1$ measure of loss is preferred.
Finally, note that a quick-and-easy (albeit suboptimal) version of the above is to
use the (supoptimal) predictor  $\hat Y_{\underline t_{k+1}}\simeq \hat \mu (\underline t_{k+1})$
and base $PRESS(b) $ or  $PRESAR(b)$ on this approximation.
\tb{For the problem of selecting $h_0$ in the case of model-free point predictors,}  as in  \cite{Politis2013},  our final choice is  $h_0=h^2$ where   $h=b/n$.
Note that an initial choice of $h_0$ 
(\tb{needed to perform} uniformization and cross-validation to determine the optimal bandwidth $b$) 
can be set by any  plug-in rule; the  effect of choosing an initial value of $h_0$ 
 is minimal.
% available in standard statistical software such as R.
% \label{RF.re.bandwidthCV}
% \end{Remark}

%%%%%%%%%%%%%%%%%%%%%%%%%%%%%%%%%%%%%%%%%%%%%%%
\section{Model-Free vs. Model-Based Inference: % for locally stationary series:
 \tr {empirical comparisons}}
\label{RF.Numerical}

The performance of the Model-Free and Model-Based \tb{predictors} described above are empirically compared using 
% both 
simulated \tb{and real-life} data
%and real-life datasets 
based on point prediction. 
%and also calculation of prediction intervals.  
The Model-Based local constant and local linear methods are denoted as \tb{MB-LC} and \tb{MB-LL} respectively. \tb{Model-Based predictors MB-LC and MB-LL are described in Section \ref{RF.Model-based inference}}. \tbl{The Model-Free methods using local constant, local linear (Hansen) and local linear (Monotone) 
% using the flat-top tapered covariance estimator 
are denoted as MF-LC, MF-LLH, MF-LLM.} Model-Free predictors are described in Section \ref{RF.Model-free inference}. Point prediction performance as indicated by %Bias and 
Mean Squared Error (MSE) are used to compare the estimators.
% \tbl{Model-Free methods using local constant, local linear (Hansen) and local linear (Monotone) \tbl{using the covariance estimator obtained from fitting a causal AR(p) model} are denoted as MF-LC-ARMA, MF-LLH-ARMA, MF-LLM-ARMA. Model-Free predictors are described in Section \ref{NSTS.Model-free inference}. The covariance estimators using the flat-top tapered kernel and fitting an AR(p) model are discussed in Section \ref{RF.seq.whiteningtransformation}.} Results are also shown for the LMF counterparts of these methods which are denoted as LMF-LC, LMF-LLH, LMF-LLM \tbl{and LMF-LC-ARMA, LMF-LLH-ARMA, LMF-LLM-ARMA respectively}. Results for all methods are given for both fitted (F) and predictive (P) residuals.
%Following metrics are used to compare the estimators:
%}
%\begin{enumerate}
%\item
%\tr {
%Point prediction performance as indicated by Bias and Mean Squared Error (MSE) on simulated and real-life datasets
%using all Model-Based and Model-Free methods listed above.
%}
%\item
%\tr {
%Bootstrap performance as indicated by coverage probability (CVR), mean length of prediction intervals and standard
%deviation (sd) of length of prediction intervals. All prediction interval metrics given in the following tables have been
%generated \tb{based on} a nominal coverage of $90\%$.
%}
%\end{enumerate}

\subsection{Simulation: Additive model with stationary 2-D AR errors}

Let a random field be generated using the 2-D AR process as below:
\begin{equation}
y(t_1, t_2) = 0.25 y_{t_1-1,t_2-1} + 0.2y_{t_1-1,t_2+1} - 0.05y_{t_1-2,t_2} + v(t_1, t_2)
\end{equation}
Let this field be generated over the region defined by $0 \leq t_1 \leq n_1 \ \& \ 0 \leq t_2 \leq n_2$ where $n_1=101, n_2=101$. The NSHP limits are set from $(101,101)$ to $(50,50)$, \tb{this defines the region $E_{\underline t, \underline n}$ as shown in  Figure \ref{NSHP_pred}}. The data $Y_{\underline t}$ is generated using the additive model in  eq.~\eqref{RF.eq.model  homo} with trend specified as $\mu(\underline t) = \mu(t_1, t_2) = \sin (4\pi \frac{t_2-1}{n_2-1})$ where $0 \leq t_1 \leq n_1 \ \& \ 0 \leq t_2 \leq n_2$. Here $v(t_1, t_2)$ are i.i.d. $N(0,\tau^2)$ where $\tau=0.1$.  Let $t_1=50, t_2=50$ where point prediction is performed. Bandwidths for estimating the trend are calculated using cross-validation for both Model-Based and Model-Free cases described in Section \ref{RF.cross-validation}.
%\tr{Data $Y_i$ for $t =1, \ldots, 1000$ were simulated as per model \eqref{RF.eq.model homo}
%with trend as in eq.~\eqref{RF.eq.qs},
%i.e., $\mu (t)= \mu_{_{[0,1]}} (a_t)$ 
%with $a_t= (t-1)/n$ and $\mu_{_{[0,1]}} (x)= \sin (2\pi x)$. 
%The series $W_t$ is constructed  via 
%an AR(5)   model  driven by errors $ V_t$ that 
%are i.i.d.~$N(0, \tau^2);$ with $\tau=0.14$.  The AR(5)
% coefficients
% are set to 0.5, 0.1, 0.1, 0.1, 0.1.
%Sample size $n$ is set to $100$. Point prediction and prediction intervals
% are measured for boundary point $t_1=50, t_2=49$. 
% 
% Bandwidths for estimating the trend are calculated using the cross-validation techniques for Model-Based and Model-Free cases described in Sections \ref{RF.sec.trend} and \ref{RF.sec.MF.cv} respectively.}

\tr {Results for point prediction using mean square error (MSE) over all MB and MF methods are shown 
% for a range of bandwidths $b$ (see Sections \ref{RF.sec.trend} and \ref{RF.sec.KEUT}) 
in Table \ref{pp_2D_AR}. A total of 100 realizations of the dataset were used for measuring point prediction performance.}  \tb{From this table it can be seen that MB-LL is the best point predictor. This is expected since the data was generated by a 2D AR model which is the same used in MB-LL prediction. In addition the estimation is performed at the boundary of the random field with a strong linear trend as shown in Figure \ref{NSHP_linear_trend} where LL regression is expected to perform the best. In addition it can be observed that MF-LLM performs the best among all MF point predictors and approaches the performance of MB-LL. This shows that monotonicity correction in the LLM distribution estimator has minimal effect on the center of the distribution that is used for point prediction.}

{\begin{figure}[!t]
  \centering
  \includegraphics[width=3.5in, height=3.0in]{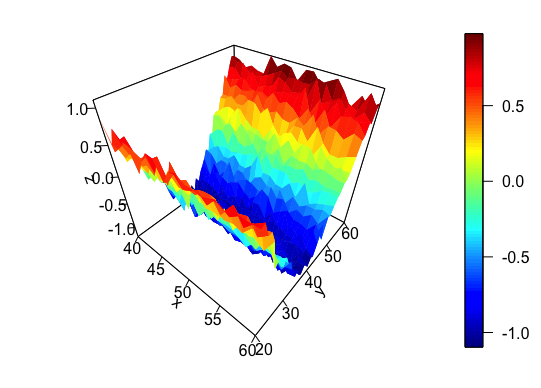}
  \caption{Linear trend for NSHP where prediction is performed  (50, 50)}
  \label{NSHP_linear_trend}
\end{figure}}

\bigskip

\begin{table}[!htbp]
\centering
\caption{Point Prediction performance for 2-D AR dataset}
\label{pp_2D_AR}
\scalebox{1.20}{
\begin{tabular}{|c|c|c|c|cccccc}
\hline
Prediction Method  & Residual Type & MSE\\
\hline
MB-LC & P  & {1.488e-02}\\
\hline
& F  & {1.520e-02}\\
\hline
MB-LL & P  & {1.393e-02}\\
\hline
& F  & {1.400e-02}\\
\hline
% MF-LC-ARMA & P  & {1.584e-02}\\
MF-LC & P  & {1.530e-02}\\
\hline
% & F  & {1.611e-02}\\
& F  & {1.549e-02}\\
\hline
% MF-LLH-ARMA & P  & {1.503e-02}\\
MF-LLH & P  & {1.471e-02}\\
\hline
% & F  & {1.567e-02}\\
& F  & {1.515e-02}\\
\hline
% MF-LLM-ARMA & P & \tb{\bf 1.459e-02}\\
MF-LLM & P & {1.414e-02}\\
\hline
% & F & \tb{\bf 1.489e-02}\\
& F & {1.456e-02}\\
\hline
\end{tabular}}
\end{table}

\subsection{Real-life example: CIFAR images}

The CIFAR-10 dataset \cite{krizhevsky2009cifar} is used as a real-life example to compare the model-based and model-free prediction algorithms discussed before. The original CIFAR-10 dataset consists of 60000 32 by 32 color images in 10 classes, with 6000 images per class. We pick 100 images from the class "dog" where the original images have 3 RGB (red, green, blue) channels with discrete pixel values. We pick the R (red) channel of each image, and standardize these to generate a new real-valued dataset. Our final transformed dataset has 100 32 by 32 random fields. The NSHP limits are set from $(32,32)$ to $(16,16)$, \tb{this defines the region $E_{\underline t, \underline n}$ as shown in Figure \ref{NSHP_pred}}. Rest of the image is considered as occluded and their pixel values are not available for prediction. Sample images used for prediction are shown in Figure \ref{CIFAR10_dog}. Let $t_1=16, t_2=16$ where point prediction is performed. Bandwidths for estimating the trend are calculated using cross-validation for both Model-Based and Model-Free cases described in Section \ref{RF.cross-validation}.

\tr {Results for point prediction using mean square error (MSE) over all MB and MF methods are shown 
% for a range of bandwidths $b$ (see Sections \ref{RF.sec.trend} and \ref{RF.sec.KEUT}) 
in Table \ref{pp_2D_AR}.}
\tb{From this table it can be seen that MF-LLH and MF-LLM are the best point predictors. We attribute this to the fact that the CIFAR-10 image data is not compatible with additive model as given by eq.~\eqref{RF.eq.model  homo}. It can also be seen that unlike the synthetic 2D AR dataset the two best predictors MF-LLH and MF-LLM are much closer in performance which is owing to lack of a linear trend at the point where prediction is performed. Lastly for point prediction there is a difference in performance between fitted and predictive residuals for some estimators which is not the case with the synthetic dataset discussed before. This is due to finite sample effects as the CIFAR image random field is smaller in size and we use only a part of this for our one-sided prediction.}

{\begin{figure}[!htbp]
  \centering
\includegraphics[width=2.8in, height=2.7in]{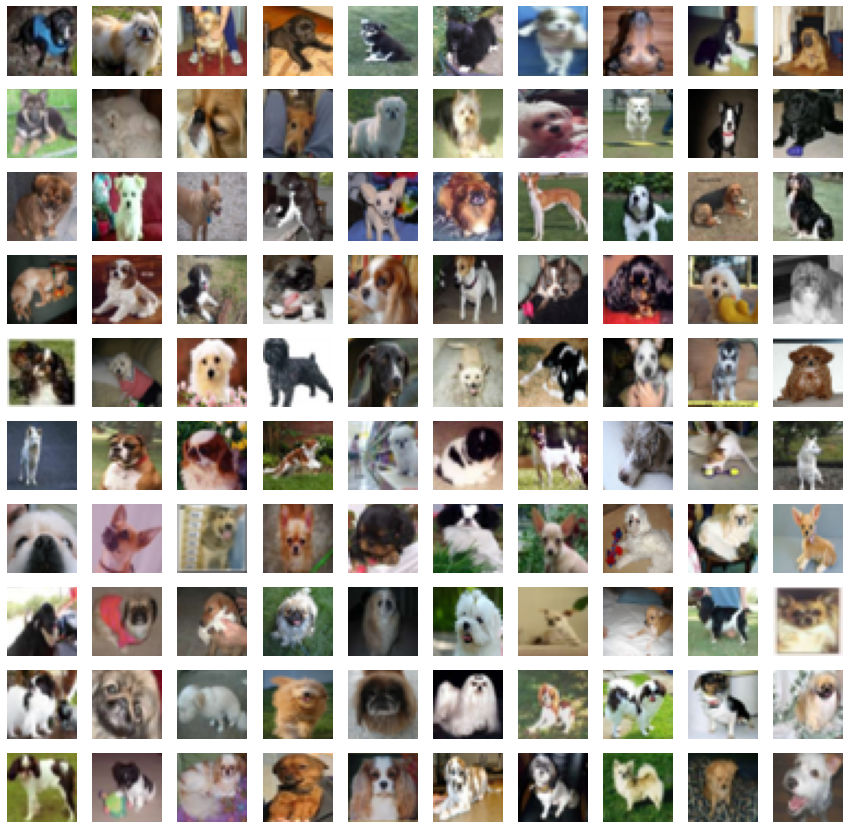}
\caption{Sample images from CIFAR-10 dataset with label dog (Note: Here full images are shown although only part of it is used for prediction.)}
\label{CIFAR10_dog}
\end{figure}}

\begin{table}[!htbp]
\centering
\caption{Point Prediction performance for CIFAR-10 dataset}
\label{pp_CIFAR}
\scalebox{1.20}{
\begin{tabular}{|c|c|c|c|cccccc}
\hline
Prediction Method  & Residual Type & MSE\\
\hline
MB-LC & P  & {1.98e-01}\\
\hline
& F  & {2.20e-01}\\
\hline
MB-LL & P  & 1.79e-01\\
\hline
& F  & 1.95e-01\\
\hline
% MF-LC-ARMA & P  & {1.584e-02}\\
MF-LC & P  & {1.79e-01}\\
\hline
% & F  & {1.611e-02}\\
& F  & {2.12e-01}\\
\hline
% MF-LLH-ARMA & P  & {1.503e-02}\\
MF-LLH & P  & {1.60e-01}\\
\hline
% & F  & {1.567e-02}\\
& F  & {1.89e-01}\\
\hline
% MF-LLM-ARMA & P & \tb{\bf 1.459e-02}\\
MF-LLM & P & {1.64e-01}\\
\hline
% & F & \tb{\bf 1.489e-02}\\
& F & {1.70e-01}\\
\hline
\end{tabular}}
\end{table}

\section{Conclusions and Future Work}
\label{RF.Numerical}

\tb{In this paper we investigate the problem of one-sided prediction over random fields that are stationary only across a limited part over their entire region of definition. For such locally stationary random fields we develop frameworks for point prediction using both a model-based approach which includes a coordinate changing trend and/or variance and also by using the model-free principle proposed by \tb{\cite{Politis2013}, \cite{politis2015model}}. We apply our algorithms to both synthetic data as well as a real-life dataset consisting of images from the CIFAR-10 dataset. In the latter case we obtain the best performance using the model-free approach and thereby demonstrate the superiority of this technique versus the model-based case where an additive model is assumed arbitrarily for purposes of prediction. In future work we plan to investigate both model-based and model-free prediction using random fields with non-uniform spacing of data as well as extending our algorithms for estimating prediction intervals.}

\vskip .1in
\clearpage
\noindent 
\tr {
{\bf Acknowledgements} \\
This research was partially supported by NSF grant DMS 19-14556. The authors would like to acknowledge the Pacific Research Platform, NSF Project ACI-1541349 and Larry Smarr (PI, Calit2 at UCSD)
for providing the computing infrastructure used in this project. 
% \tbl{Many thanks are also due to Richard Davis and Stathis Paparoditis for their helpful comments.}
}
\bibliographystyle{plain} 
\bibliography{srinjoy_stats}

\begin{thebibliography}{10}

\bibitem{brockwell2013time}
Peter~J Brockwell and Richard~A Davis.
\newblock {\em Time series: theory and methods}.
\newblock Springer, New York, second edition, 1991.

\bibitem{choi2007modeling}
ByoungSeon Choi and Dimitris~N Politis.
\newblock Modeling 2-d ar processes with various regions of support.
\newblock {\em IEEE transactions on signal processing}, 55(5):1696--1707, 2007.

\bibitem{dahlhaus2012locally}
Rainer Dahlhaus.
\newblock Locally stationary processes.
\newblock In Tata~Subba Rao et~al., editors, {\em Handbook of statistics},
  volume~30, chapter~13, pages 351--412. Elsevier, 2012.

\bibitem{dahlhaus1997fitting}
Rainer Dahlhaus et~al.
\newblock Fitting time series models to nonstationary processes.
\newblock {\em The Annals of Statistics}, 25(1):1--37, 1997.

\bibitem{das2019nonparametric}
Srinjoy Das and Dimitris~N Politis.
\newblock Nonparametric estimation of the conditional distribution at
  regression boundary points.
\newblock {\em The American Statistician}, 2019.

\bibitem{das2021predictive}
Srinjoy Das and Dimitris~N Politis.
\newblock Predictive inference for locally stationary time series with an
  application to climate data.
\newblock {\em Journal of the American Statistical Association},
  116(534):919--934, 2021.

\bibitem{dudgeon1984multidimensional}
Dan~E Dudgeon and Russell~M Mersereau.
\newblock {\em Multidimensional Digital Signal Processing Prentice-Hall Signal
  Processing Series}.
\newblock Prentice-Hall, Englewood Cliffs, NJ, 1984.

\bibitem{fan1996local}
Jianqing Fan and Irene Gijbels.
\newblock {\em Local polynomial modelling and its applications: monographs on
  statistics and applied probability}, volume~66.
\newblock CRC Press, Boca Raton, 1996.

\bibitem{fan2007nonlinear}
Jianqing Fan and Qiwei Yao.
\newblock {\em Nonlinear time series: nonparametric and parametric methods}.
\newblock Springer, New York, 2007.

\bibitem{hansen2004nonparametric}
Bruce~E Hansen.
\newblock Nonparametric estimation of smooth conditional distributions.
\newblock {\em Unpublished paper: Department of Economics, University of
  Wisconsin}, 2004.

\bibitem{hardle1992kernel}
Wolfgang H{\"a}rdle and Philippe Vieu.
\newblock Kernel regression smoothing of time series.
\newblock {\em Journal of Time Series Analysis}, 13(3):209--232, 1992.

\bibitem{kim1996bandwidth}
Tae~Yoon Kim and Dennis~D Cox.
\newblock Bandwidth selection in kernel smoothing of time series.
\newblock {\em Journal of Time Series Analysis}, 17(1):49--63, 1996.

\bibitem{krizhevsky2009cifar}
Alex Krizhevsky, Vinod Nair, and Geoffrey Hinton.
\newblock Cifar-10 and cifar-100 datasets.
\newblock {\em URl: https://www. cs. toronto. edu/kriz/cifar. html}, 6(1):1,
  2009.

\bibitem{kurisu2022nonparametric}
Daisuke Kurisu.
\newblock Nonparametric regression for locally stationary random fields under
  stochastic sampling design.
\newblock {\em Bernoulli}, 28(2):1250--1275, 2022.

\bibitem{li2007nonparametric}
Qi~Li and Jeffrey~Scott Racine.
\newblock {\em Nonparametric econometrics: theory and practice}.
\newblock Princeton University Press, Princeton, 2007.

\bibitem{Politis2013}
Dimitris~N Politis.
\newblock Model-free model-fitting and predictive distributions.
\newblock {\em Test}, 22(2):183--221, 2013.

\bibitem{politis2015model}
Dimitris~N Politis.
\newblock {\em Model-Free Prediction and Regression}.
\newblock Springer, New York, 2015.

\bibitem{priestley1965evolutionary}
Maurice~B Priestley.
\newblock Evolutionary spectra and non-stationary processes.
\newblock {\em Journal of the Royal Statistical Society. Series B
  (Methodological)}, pages 204--237, 1965.

\bibitem{priestley1988non}
Maurice~Bertram Priestley.
\newblock {\em Non-linear and non-stationary time series analysis}.
\newblock Academic Press, London, 1988.

\bibitem{samorodnitsky1994stable}
Gennady Samorodnitsky and Murad~S Taqqu.
\newblock Stable non-gaussian random processes: Stochastic models with infinite
  variance (stochastic modeling series), 1994.

\bibitem{zhou2009local}
Zhou Zhou and Wei~Biao Wu.
\newblock Local linear quantile estimation for nonstationary time series.
\newblock {\em The Annals of Statistics}, 37(5B):2696--2729, 2009.

\bibitem{zhou2010simultaneous}
Zhou Zhou and Wei~Biao Wu.
\newblock Simultaneous inference of linear models with time varying
  coefficients.
\newblock {\em Journal of the Royal Statistical Society: Series B (Statistical
  Methodology)}, 72(4):513--531, 2010.

\end{thebibliography}

\end{document}